\documentclass{aa}
\usepackage{epstopdf}
\usepackage{graphicx}
\usepackage{rotating}
\usepackage{txfonts}
\usepackage{color}
\usepackage{natbib}
\bibpunct{(}{)}{;}{a}{}{,}

\begin{document}
\newcommand{\dd}{deg$^{2}$}
\newcommand{\flux}{$\rm erg \, s^{-1} \, cm^{-2}$}
\newcommand{\LL}{$\lambda$}

   \title{The cosmological analysis of X-ray cluster surveys}
   \subtitle{IV. Testing ASpiX with template-based cosmological simulations}  
   \titlerunning{The cosmological analysis of X-ray cluster surveys.IV}   
   \authorrunning{Valotti, Pierre et al}
   \author{A. Valotti\inst{1, 2}, M. Pierre\inst{1, 2}, A. Farahi\inst{3}, A. Evrard\inst{3}, L. Faccioli\inst{1, 2}, J.-L. Sauvageot\inst{1, 2}
   N. Clerc\inst{4, 5, 6}, F. Pacaud\inst{7}        
          }
   \institute{IRFU, CEA, Universit\'e Paris-Saclay, F-91191 Gif-sur-Yvette, France 
\and   
   Universit\'e Paris Diderot, AIM, Sorbonne Paris Cit\'e, CEA, CNRS, F-91191 Gif-sur-Yvette, France
         \and
Departments of Physics and Astronomy and Michigan Center for Theoretical Physics,  University of Michigan, Ann Arbor, MI USA 
\and
Max Planck Institut f\"ur Extraterrestrische Physik, Giessenbachstrasse 1, 85748 Garching bei M\"unchen, Germany 
\and
 CNRS, IRAP; 9 Av. colonel Roche,  F-31028 Toulouse cedex 4, France
 \and
Université de Toulouse, UPS-OMP; IRAP; Toulouse, France
\and
Argelander Institut f\"ur Astronomie, Universit\"at Bonn, 53121 Bonn, Germany            
             }

   \date{accepted for publication in A\& A}  
   
  \abstract
{This paper is the fourth of a series evaluating the ASpiX cosmological method,  based on X-ray diagrams, which are constructed from simple cluster observable quantities, namely: countrate (CR), hardness ration (HR), core radius ($r_c$) and redshift.}
   {Following extensive tests on analytical toy-catalogues (paper III), we present the results of  a more realistic study over a 711 \dd\ template-based  maps derived from a cosmological simulation.   }
   {Dark matter halos from the Aardvark simulation have been ascribed luminosities, temperatures and core-radii, using local scaling relations and assuming self-similar evolution. The predicted X-ray  sky-maps are converted into XMM event lists, using a detailed instrumental simulator. The XXL pipeline run on the resulting sky images, produces an `observed' cluster catalogue over which the tests have been performed. This allowed us to investigate the relative power of various combinations  of the count-rate, hardness-ratio, apparent-size and redshift information. Two fitting methods were used : a traditional MCMC approach and a simple minimisation procedure  (Amoeba) whose mean uncertainties are a posteriori evaluated by means of synthetic catalogues. The results were analysed and compared to the predictions from the Fisher analysis.    }
   {For this particular catalogue realisation, assuming that the scaling relations are perfectly known, the CR-HR combination gives  $\sigma_8$ and $\Omega_m$ at the 10\% level, while CR-HR-$r_c$-z improves this to $\leq$ 3\%. Adding a second hardness ratio improves the results from the CR-HR1-$r_c$ combination, but to a lesser extent than when adding the redshift information.  When all coefficients of the M-T relation (including scatter) are also fitted, the cosmological parameters are constrained to within 5-10 \%, and larger for the M-T coefficients (up to a factor of two for the scatter). 
The errors returned by the MCMC, those by Amoeba and the Fisher Analysis predictions are in most of the cases in excellent agreement and always within a factor of two.
We also study the impact of the scatter of the M-Rc relation on the number of detected clusters: for the cluster typical sizes usually assumed, the larger the scatter, the lower the number of detected objects. 
}
   {The present study  confirms and extends the trends outlined in our previous analyses, namely the power of X-ray observable diagrams to successfully and easily fit at the same time, the cosmological parameters, cluster physics and the survey selection, by involving all detected clusters.  The accuracy levels quoted here should not be considered as definitive: a number of simplifying hypotheses were made for the testing purpose, but this should affect in the same way any method. The next publication will consider in greater detail the impact of cluster shapes (selection and measurements) and of cluster physics in the final error budget by means of hydrodynamical simulations.  }

   \maketitle
%

\section{Introduction}

Clusters of galaxies constitute one of the low-redshift cosmological probes complementing early universe measurements from the cosmic microwave background (CMB). Since cluster number counts are both sensitive to the geometry of the universe and to the growth of structure, related statistics provide, in theory, key cosmological information. But because of the many uncertainties impinging on cluster mass determination, the reliability of the cluster route has been time after time questioned. In the past few years however, there is growing evidence that independent  cosmological analyses based on structure growth at low-z favour a lower $\sigma_{8}$ than the most recent Planck CMB studies \citep{planck2015XXIV,pacaud16}. In other words, we find fewer clusters of a given mass than the CMB cosmology predicts, given our current knowledge of cluster physics as coded in the mass-observable relations. Cluster are thus expected to provide a critical contribution to the up-coming  extensive dark energy studies.

Cluster cosmology requires jointly modelling the physical parameters describing the evolution of the intra-cluster medium  along with the impact of selection procedure. While the first self-consistent methods have been been following a backward modelling of the recovered cosmology-dependent, mass function \citep[e.g.][]{vikhlinin09}, more recent studies moved to a forward approach whose likelihood includes physical quantities such as luminosity, temperature or gas fraction \citep[e.g.][]{mantz14,mantz15}. Cluster number counts from Sunyaev-Z'eldovich surveys are routinely modelled in terms  of the signal-to-noise-ratio or of the Compton parameter of the  detections, which can be related to the cluster mass via scaling relations (X-ray, lensing, velocities) 
\citep[e.g.][]{vanderlinde10,hassel13,benson13,bocquet15,planck2015XXIV}.
In this context, we are developing a cosmological analysis method (ASpiX) based on X-ray cluster number counts that does not explicitly rely either on cluster mass determinations or on physical quantities. It consists in the  modelling of the multi-dimensional distribution of a set of directly measurable X-ray clusters quantities, namely:  count-rates, colors, apparent size, which are all cosmology-independent. This method is particularly suited to rather shallow survey-type data, when the number of collected X-ray photons is too low to enable detailed spectral and morphological analyses.  Thanks to its modularity, the ASpiX method considerably eases the process by simultaneously fitting in the observed parameter space, the effect of cosmology, selection and cluster physics. Depending on the volume surveyed, i.e. on the number of clusters involved in the analysis, the number of parameters that may be fitted can increase from a few to  15 or more, including in particular scatter and evolution in the scaling relations. This method cannot rival approaches including deep pointed X-ray observations along with ancillary data from other wavebands and, fundamentally, faces the same uncertainties as to the observable-mass transformation. However, it allows the inclusion of the vast majority of the detected clusters even when only a few tens of photons are available. Furthermore, when cosmological simulations are produced at a significantly high rate, the method will allow us to totally bypass any mass estimate or scaling-relation related formalism; instead, it will solely rely on the simulations by comparing the observed and simulated parameter distributions \citep{pierre17a}. In the end, neither assumptions based on the hydrostatic equilibrium nor any modelling of the mass function will be necessary.

This paper is the fourth of a series aiming at an in-depth characterisation of the ASpiX method, with the ultimate goal of applying it to the current large X-ray cluster surveys. Our philosophy is to address a few specific issues per article: Paper I \citep{clerc12a} laid out the principle of the method. In paper II \citep{clerc12b}, we applied ASpiX on a  347 cluster sample drawn from the XMM archive, assuming fixed scaling relations and we provided predictions for the eRosita survey. Paper III \citep{pierre17a} was devoted to the systematic exploration of the ASpiX behaviour by means of analytical cluster toy-catalogues:  impact of the resolution of the observed parameter-space, particular role of the cluster apparent-size information, optimisation of a fast minimisation procedure (Amoeba), error estimates, search for possible degeneracies between cosmology and cluster physics in the various parameter-space representations.  In this fourth paper, we pursue our evaluation of the ASpiX method, now in almost real-world conditions, i.e. by analysing synthetic X-ray images: we assume a more realistic error model for the observable quantities, we study the effect of using a second hardness ratio and of scatter in the mass-size relation, we detect the impact of projection effects in the selection function, we compare the Amoeba-dependent minimisation and error estimates with a standard MCMC fitting.  The next, and last, validation article will quantitatively evaluate the systematic errors by means of hydrodynamical simulations. \\
The synthetic images in the present paper are produced by applying emission template forms to halo populations realised in N-body simulations. The images are transformed into XMM observations taking into account all instrumental and background effects. The simulated images are in turn processed as regular XMM pointings and the detected clusters of galaxies are selected following a well-defined procedure. Finally, the ASpiX method is run on the selected sample and  the derived cosmological parameters are compared to those of the input numerical simulations.

The paper is organised as follows. The next Section recalls the basis of the ASpiX method. Section 3  describes the numerical simulations and the mapping of the X-ray properties onto the dark matter halos. Section 4 explains the transformation of the simulated X-ray sky maps into XMM images. The reduction of the XMM images along with the production of the resulting cluster catalogues is presented in Section 5. The results of the cosmological analysis of the cluster catalogues are given in Section 6. In Section 7, we analyse the results and the impact of particular cluster parameters. Last Section draws the conclusions and outlines the future steps. The cosmological model adopted for this test-case study is presented in Sec. \ref{AardvarkSim}  and summarised in Table \ref{allparam}.

\section{The ASpiX method}
In all what follows, we assume that the X-ray observations are performed with an XMM survey, but the principle can be easily translated to any X-ray telescope (e.g. Chandra, eRosita).
\subsection{Modelling the cluster population in the XOD}
\label{aspix-m}
\begin{figure}
   \centering
      \includegraphics[width=8cm]{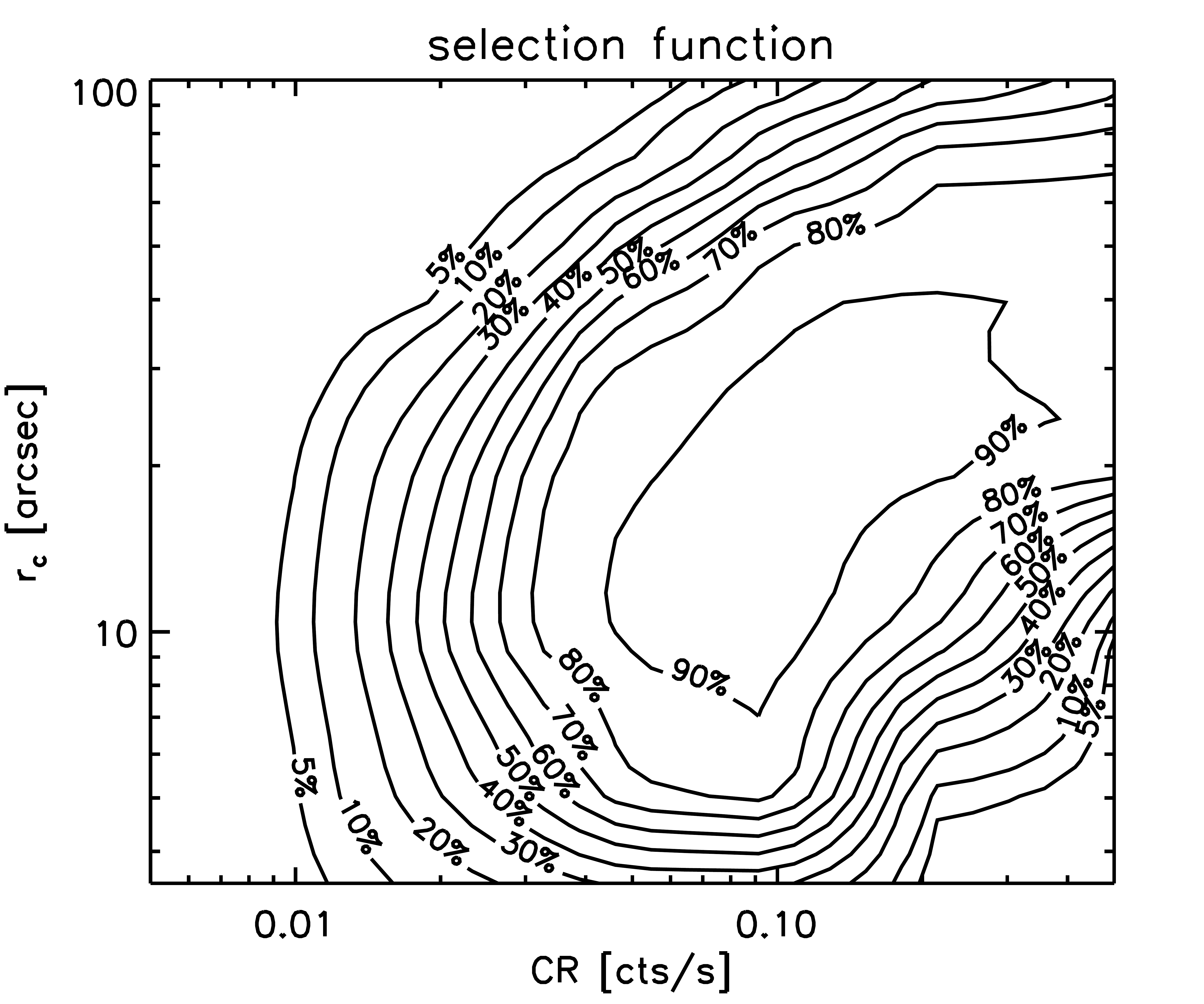}
   \caption{Selection function adopted for the present study.   The probability to detect a cluster as C1 is given by the isocontours as a function of count-rate and core-radius. This map has been derived from extensive XMM image simulations and the two axes stand for the true (input) cluster parameters; it is thus only valid for the conditions under which the simulations were run (XMM exposure time of 10ks and background).
}
 \label{selfunc}
    \end{figure}

\begin{figure}
   \centering
\includegraphics[width=8cm]{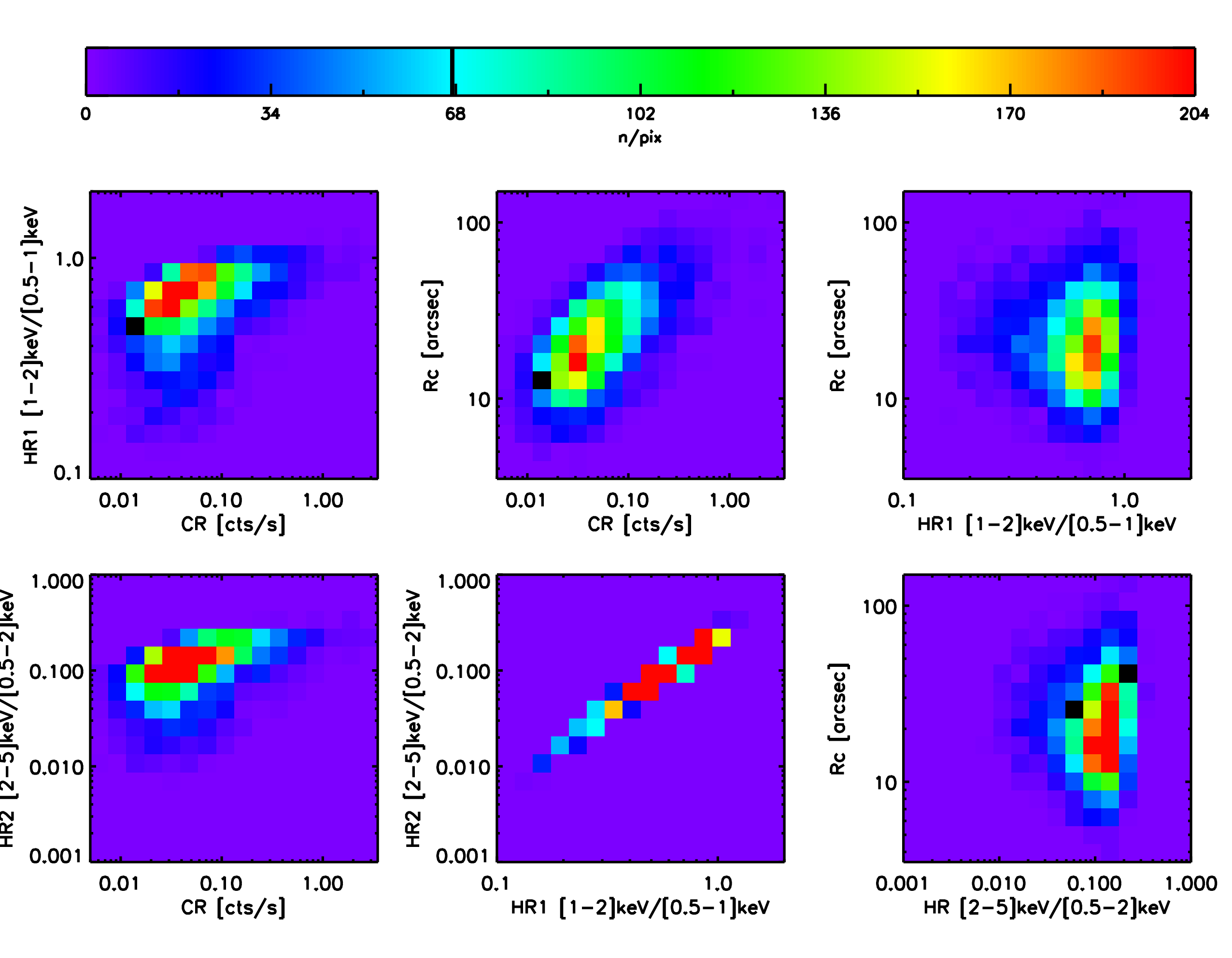}
   \caption{X-ray Observable Diagram computed for a 700 \dd\ cluster survey, observed with 10 ks XMM exposures. The six panels show the 2D  projections of the distribution of the four cluster parameters involved in the present study:  count-rate (CR) in [0.5-2] keV, hardness ratio HR$_1$ ([1-2]/[0.5-1] keV), HR$_2$ ([2-5]/[0.5-2]) keV, angular cluster size $r_{c}$. The diagrams are integrated over the $0<z<2$ range, but this fifth dimension can be uncompressed if redshifts are available, which significantly increases the cosmological constraining power of the ASpiX method. Error measurements are not implemented in this example. }
              \label{XODex}
    \end{figure}

The principle of the ASpiX method consists in the fitting of a multi-dimensional distribution  of X-ray observable parameters drawn from a selected cluster population. 
The so-called X-ray Observable Diagrams (XOD) involve part or all of the following parameters: instrumental count-rates (CR) and hardness ratios (HR) in well-specified bands, a measurement of the cluster angular size ($r_{c}$) and the redshift (z, assumed to be measured by optical ground-based observations). The method relies on the fact that the cluster mass information as a function of redshift, i.e. our link to cosmology, is encoded in the combination of these parameters. 
In practice, we model XODs by assuming:
\begin{itemize}
\item[•] A cosmological model 
\item[•] A cluster mass function
\item[•] X-ray cluster scaling relations, including scatter and evolution ($M-T,~ L-T~, M-Rc$)
\item[•] A plasma code to transform luminosities into fluxes as a function of temperature, abundances and redshift.
\item[•] A model for the X-ray cluster emission profile
\item[•] The XMM response to convert fluxes into count-rates and a PSF model to convolve the cluster profiles
\item[•] Total area and XMM exposure time for the survey in question
\item[•] An error measurement model and a realistic cluster selection function for a given detection pipeline, calibrated using extensive simulations 
\end{itemize}
The numerical ingredients of the model  are given in Sec. \ref{simulations}. We stress that, because clusters are extended sources, the cluster selection is performed in a two-dimensional parameter space, [CR, $r_{c}$], which is equivalent to the physical [flux, apparent size] plane. The adopted selection function is analogous to that of the XXL survey \citep{pierre16} and is given in Fig. \ref{selfunc}; we refer the reader to paper III for a detailed description. 
An example of a 4-dimensional XOD is displayed in Figure  \ref{XODex}.

\subsection{Fitting the X-ray Observable Diagrams}
\label{aspix-f}
The cosmological analysis of an X-ray cluster survey with ASpiX consists in finding the combination of the cosmological + cluster physics parameters that best fits the observed XOD. This is done by varying the parameter values of the model. The number of parameters that can be simultaneously fitted  depends on the survey area and on the measurement accuracy. An obvious choice for the minimisation procedure is the MCMC approach and was the method adopted for the fitting of the XOD obtained from the XMM archive in paper II. The computer time, however, increases very rapidly as a function of the number of free parameters when 4-dimensional XOD are considered and so, becomes prohibitive for the current testing phase. We thus favour a simple minimisation procedure (Amoeba \citep{nelder65}), that allows us to identify in relatively short time the most likely solution. The drawback is that this procedure does not provide uncertainties on the best fitting parameters. However, as shown in paper III, reliable error estimates can be obtained by averaging the output from at least ten different toy-catalogue realisations, drawn for that purpose. In this paper, we run both approaches in parallel  to test the consistency of the results. As a complement, we give the predictions from the Fisher analysis: although, strictly speaking, only valid for Gaussian posterior distributions, this analysis provides us with a potentially quick tool to perform cosmological predictions; it is thus useful to estimate how close these predictions are to the results that we ought to achieve. 

\subsection{General settings of the present study}
\label{aspix-s}
Keeping in mind the goal of the present paper, that is testing ASpiX on synthetic surveys from cosmological simulations, we use:
\begin{itemize}
\item[•] The Aardvark simulations (Sec. \ref{AardvarkSim}) that provide us with with a projected light-cone of dark matter halos over a volume of some 700 \dd\ out to a redshift of  2. 
\item[•] The cluster physics parameters listed in Sec. \ref{ingredients} with two options for the cluster emissivity profiles to map the X-ray properties of the dark halos (Table \ref{configx}).
\item[•] XMM individual observing times of 10 ks 
\item[•] The detection pipeline and the C1 cluster selection function  that are routinely used for the XXL survey.
\item[•] Either the simple Amoeba minimisation procedure or a  MCMC analysis
\end{itemize}
In the framework of testing the ASpiX method in increasingly realistic conditions, the most significant upgrade with respect to paper III is the fact that cluster detection is now performed on maps having a more realistic distribution of source halos than the toy-catalogues. We also take the opportunity to investigate the effect of cluster core radii and of scatter in the M-Rc relation on the number of detected clusters, hence on the cosmology. We consider a more realistic model for the measurement errors. Moreover, we introduce a second hardness ratio, $HR_{2}=  [2-5]/[0.5-2]$ keV in addition to $HR_{1}=  [1-2]/[0.5-1]$ keV.\\
We describe in the next three sections the production of simulated XMM cluster catalogues, that  constitute the input of our cosmological analysis. 

\begin{table}[t]
\begin{center}
\begin{tabular}{c c }
\hline
\hline
$\Omega_m$&0.23\\
$\Omega_{\Lambda}$&1-$\Omega_m$\\
$\sigma_8$ & 0.83\\
$w_0$ &-1\\
$h$ & 0.73\\
\hline
$C^{MT}$ & 0.46\\
$\alpha_{MT}$ & 1.49\\
$\gamma_{MT}$ & 0.0\\
$\sigma_{\ln M|T}$& 0.1\\
\hline
$C^{LT}$ & 0.40\\ 
$\alpha_{LT}$ & 2.89\\
$\gamma_{LT}$ & 0.0\\
$\sigma_{\ln L|T}$ & 0.27\\
\hline
$x_{c}$& 0.24\\
$\sigma_{\ln R_c|R_{500c}}$ & 0.5 \\
\hline
\end{tabular}
\caption{Main cosmological and cluster physics parameters used in this study. The cluster scaling relations read: $L \propto 10^{C^{LT}}T^{\alpha_{LT}} E(z) (1+z)^{\gamma_{LT}}$; $M \propto 10^{C^{MT}}T^{\alpha_{MT}} E(z)^{-1} (1+z)^{\gamma_{MT}}$ - see Sec. \ref{scalrel}}
\label{allparam}
\end{center}
\end{table}
\noindent

\section{Large-scale X-ray emissivity maps of the intra-cluster medium}
We present in this section the production of 25 \dd\ emissivity maps of the intra-cluster medium (ICM) using template-based N-body simulations. 
\label{simulations}
\subsection{The Aardvark simulations}
\label{AardvarkSim}
\begin{figure*}[ht]
\centering
\centerline{\includegraphics[width=6cm]{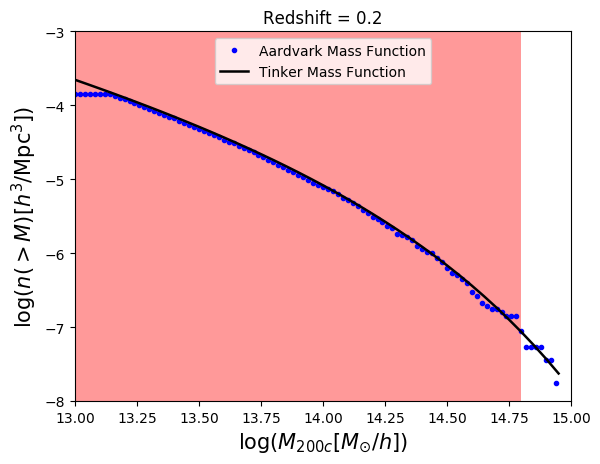},\includegraphics[width=6cm]{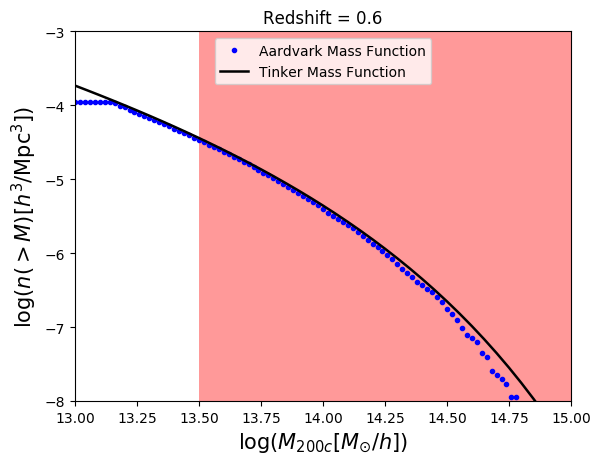},\includegraphics[width=6cm]{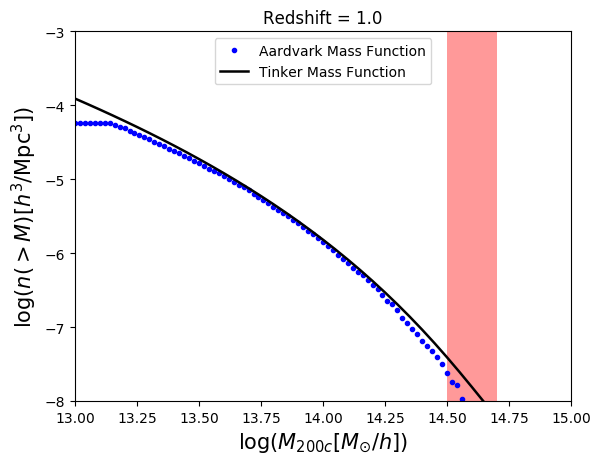} }
\caption{Cumulative dark matter halo number  density as a function of mass at different epochs. Blue dots:   Aardvark simulations. The pink areas show the mass range  encompassed by the C1 selection. The mass scale of $10^{13.2} M_\odot$ represents the halo mass resolution limit of the simulations}
\label{aamasscmp}
\end{figure*}

We employ N-body simulations produced on XSEDE resources \citep{erickson13} with a lightweight version of the Gadget code developed for the Millennium Simulation \citep{springel05}. Three simulations, of $1.05$, $2.6$ and $4.0 $ Gpc$^3$/h  volumes, are used to produce a sky survey realization covering 10,000 deg$^2$ that resolves all halos above $10^{13.2} M_{\odot}$ within $z \le 2$.  The resulting sky catalogue is built by concatenating continuous light-cone output segments from the three different N-body volumes using the method described in \citep{evrard02}.
The smallest volume maps $z < 0.35$, the intermediate maps $0.35 \le z < 1.1$ and the largest volume covers $1.1 \le z < 2$.
The simulations employ $2048^3$ particles, except for the $1.0$ Gpc$^3$/h volume which uses $1400^3$, and corresponding particle masses are $0.27$, $1.3$ and  $4.8 \times 10^{11}M_{\odot}/h $.
The Aardvark suite assumes a $\Lambda$CDM cosmology with cosmological parameters: 
$\Omega_m = 0.23$, $\Omega_{\Lambda} = 0.77$, $\Omega_{b} = 0.047$, $\sigma_8 = 0.83$, $h = 0.73$, and $n_s = 1.0$.
The Rockstar algorithm is used for halo finding \citep{behroozi13}.  
We refer to this suite of runs as the Aardvark simulation (for more detail see \cite{farahi16}).  Fig. \ref{aamasscmp} compares the mass function of the Aardvark halos to Tinker's \citep{tinker08}, which is used in our analytical fit model.  

\noindent

\subsection{The X-ray properties of clusters with a template approach}

\label{scalerel}

\begin{figure}
\centerline{\includegraphics[width=9.5cm]{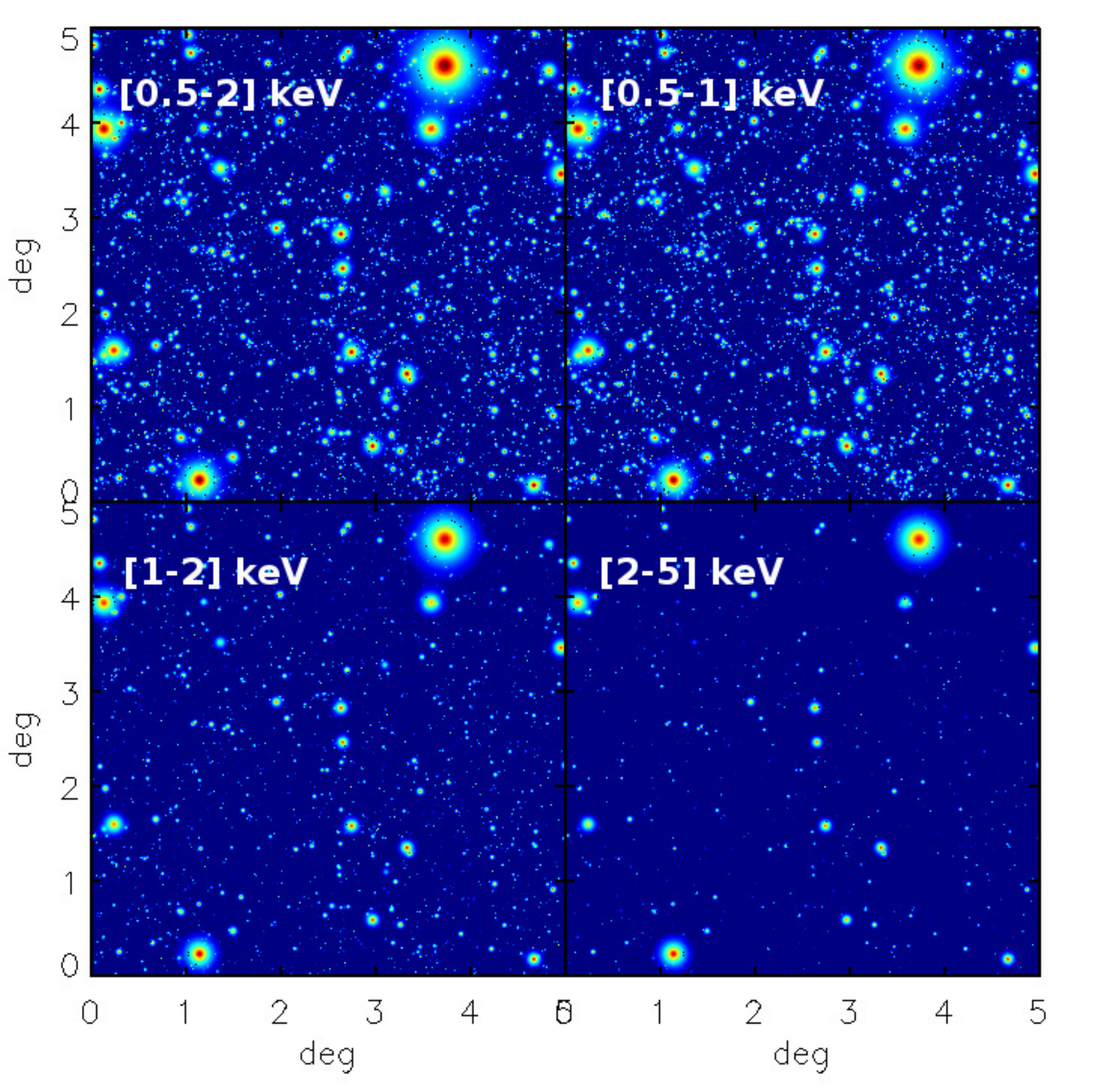}}
\caption{Typical ICM emissivity maps from the Aardvark simulations. The panels show a 25 \dd\ field  in the four bands of interest for the current study, namely:  [0.5-2] keV, [0.5-1] keV, [1-2] keV and [2-5] keV. No background, instrumental effects and AGN are added. Cluster detection is performed in the [0.5-2] keV band}
\label{maps}
\end{figure}

Starting from the Aardvark dark matter halo population, we map the ICM properties  using a standard population model \citep{evrard14}.
These models are motivated  by theoretical arguments \citep{kaiser86} and they rely on empirical data reflecting our current knowledge of the baryonic component, which mostly pertains to the high end of the mass function. We  extrapolate these models to lower mass halos in order to include  galaxy groups, that constitute the bulk of the population encompassed by our selection function (Fig.\ref{aamasscmp}). We follow the traditional modelling of the cluster gas mass and X-ray properties by means of power-law scaling relations and assume log-normal covariance. These assumptions are supported by numerous observations, theoretical arguments and simulation findings \citep[e.g.][]{kaiser86,kravtsov06,lebrun14,mantz16,mccarthy17}. \\
Practically, we begin with the mass, redshift, and sky location of dark matter halos in the Aardvark simulation. Then, we use scaling relations to infer  the mean gas temperature  and  bolometric luminosity. By means of the APEC plasma code, we deduce the X-ray fluxes in the bands of interest for the present study, namely :  [0.5-1.0], [1.0-2.0], [0.5-2.0] and   [0.2-5.0] keV. The halo X-ray surface brightness profiles are assumed to follow a $\beta$ model.  This allows us to produce theoretical X-ray emissivity maps, of which we show an  example in Fig. \ref{maps}.
At this stage, we stress that only the [0.5-2] keV map is used in the current study: this is the band where the source detection is performed; fluxes in the other bands are analytically derived (Sec. \ref{correl}).

\subsection{Ingredients of the cluster modelling}
\label{ingredients} 
The particular ingredients of the cluster X-ray mapping are given in the following paragraphs

\subsubsection{Mass overdensity}
The Tinker mass function is computed at an overdensity of $\delta \rho =200$ (mean density) and transformed into a function of $M_{200c}$ (critical) by means of a NFW profile and a concentration-mass relation \citep{navarro97,hu03,bullock01}. To switch from the $M_{200c}$ parameter of the Aardvark simulations (and the $M_{200c}$ Tinker mass function) to the $M_{500c}$ value, we assume the empirical relation:
$M_{500c}/M_{200c} = 0.714$,   
following \cite{lin03}.
This is used for the scaling relation of $R_{c}$ (see below).

\subsubsection{X-ray luminosity and temperature}
\label{scalrel}
We model the cluster scaling relations as power laws following self-similar evolution:
\begin{equation}
\frac{M_{200c}}{10^{14}h^{-1}M_{\odot}} = 10^{0.46} \left( \frac{T_x}{4keV} \right)^{1.49} E(z)^{-1}
\label{mteq}
\end{equation} 

\citep{arnaud05}
\begin{equation}
 \frac{L_{Xbol}}{10^{44}erg/s} =
 10^{0.40} \left( \frac{T_x}{4keV}\right) ^{2.89}E(z).
 \label{lteq}
\end{equation} 
\citep{pratt09}

In both relations, we allow for intrinsic scatter $\sigma_{lnT|M}$ and  $\sigma_{lnL|T}$. Scatter in both measures, which reflects the various merging histories and relaxation states of the halos,  are assumed here to be uncorrelated and independent of redshift and mass.  We take 0.1 and 0.27 for $\sigma_{lnT|M}$ and  $\sigma_{lnL|T}$ respectively.

\subsubsection{Cluster profiles}
Halos are assumed to be spherically symmetric:
$R_{500c} =  3/4\pi\times (M_{500c}/\rho_{500c}(z))^{1/3}$.
Cluster surface brightness profiles are modelled with a simple standard $\beta$-profile \citep{cavaliere76}. 
\begin{equation}
S(r) = S_0 \Bigg [ 1+\Bigg ( \displaystyle \frac{r}{r_{c}}\Bigg)^2 \Bigg]^{-3\beta +1/2}
    \label{surf}
\end{equation}
where $r$ and $r_{c}$ are the projected profile coordinate and the core radius. The cluster angular size ($r_{c}$) is given by $r_{c}[\text{arcsec}] \propto R_{c}[\text{Mpc}]/Da(z)[\text{Mpc}]$, where $Da$ is the angular distance diameter. We further relate the cluster core radius to the cluster size by :
$R_{c}=x_{c} \times R_{500} $, which yields
\begin{equation}
M_{500c} = \frac{4 \pi}{3}\left(\frac{R_{c}}{x_c}\right)^3\times\rho_{500c}(z)
\end{equation}
We analysed the OWLS hydrodynamical simulations \citep{lebrun14} to obtain a plausible mean estimate, given the redshift and mass ranges pertaining to the present study. Assuming a $\beta$  of 2/3, we find a mean value of 0.24 (a value also observationally found in paper II) for $x_{c}$ with a $\sigma_{\ln R_{c}|R_{500c}}$ of 0.5.
In the present paper, we shall stick to a constant 0.24 value, allowing or not for scatter. We thus analyse two X-ray mappings of the Aardvark halos as summarized in Table \ref{configx}. In the final discussion, we explore the impact of other  $x_{c}$ and scatter values on the number of detected clusters. The cluster physics parameters assumed for this study are summarised in Table \ref{allparam}.

\begin{table}[t]
       \caption{Adopted values for the X-ray emission profile of the Aardvark simulated halos.}
       \centerline{\\}
\centerline{\renewcommand{\arraystretch}{1.2}
\begin{tabular}{c c| c }
\multicolumn{2}{c}{Configuration ID} &  $\sigma_{\ln R_{c} | R_{500c}}$ \\        
\hline
\hline
$\beta=2/3$  & $x_{c}=0.24$\\
\multicolumn{2}{c|}{\textbf{B0}} &  \textbf{0} \\
\multicolumn{2}{c|}{\textbf{B0.5}} & \textbf{0.5} \\
\hline
\end{tabular}}
\label{configx}
\end{table}

\begin{figure}
\centering
\centerline{\includegraphics[width=9.5cm]{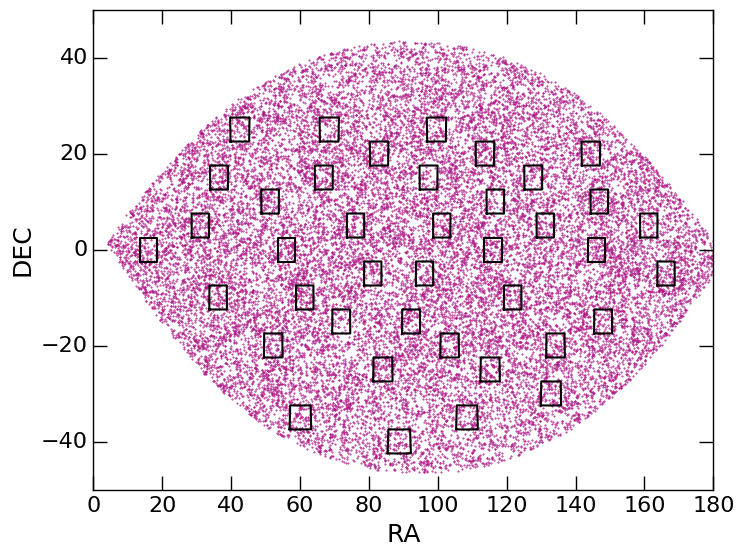}}
\caption{Sky maps of the 39 regions extracted from the Aardvark simulations. Each square covers 25 deg$^2$. Magenta dots show all halos with a mass larger then 10$^{14}$ M$_{\odot}$. The simulation depth is 0<z<2.}
\label{tiles}
\end{figure}

\subsection{Photon maps}
\label{photonmap}
\begin{figure}
\centerline{\includegraphics[width=9.5cm]{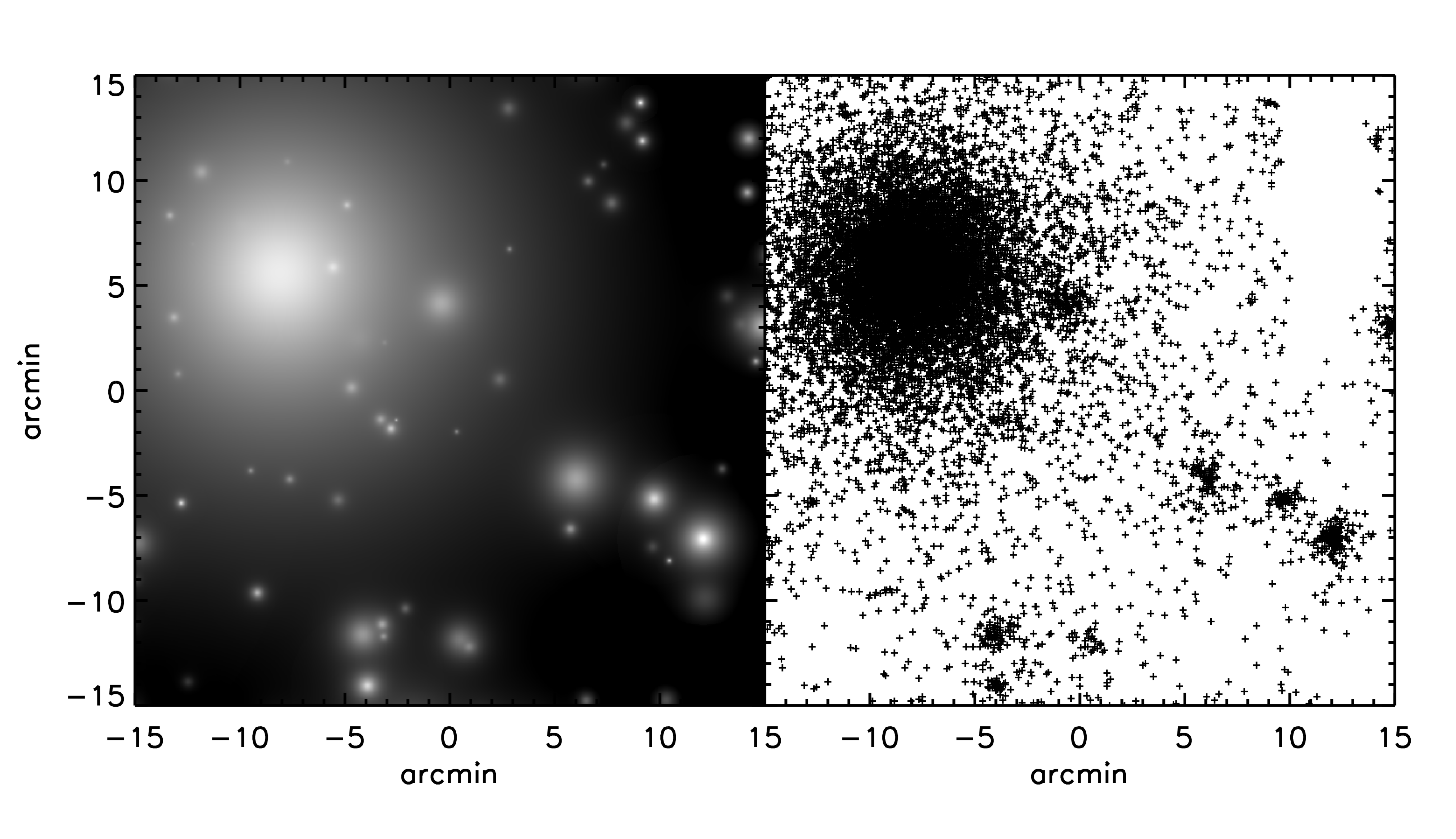}}
\caption{Left: Example of an ICM X-ray emissivity map in the [0.5-2] keV band. Right: corresponding photon image assuming a 10ks exposure and a collecting area of 1000 cm$^2$. The images are 30'$\times$30' and have a pixel size of 2.5".}
\label{photconv}
\end{figure}

We extract from the Aardvark simulated sky 39 subregions of 25 \dd\ each, randomly distributed  and sufficiently distant from each other, so that the effects of covariance between the samples are negligible when considering the total area of 975 \dd (Fig. \ref{tiles}). The size of the individual regions has been chosen such as to match that of the two XXL fields. The large number of subregions provides us with a useful handle  to estimate the uncertainty on the cosmological parameters via ASpIX in its Amoeba implementation. \\
Bolometric luminosity, temperature and X-ray profile are ascribed to each halo characterised by its mass and redshift, following the prescriptions of Sec. \ref{ingredients}. From this, we map the X-ray halo emissivity in our detection band ([0.5-2] keV) by means of the APEC X-ray plasma code (\cite{clerc12a}, \cite{pierre17a}) following the atomic densities reported by \citep{grevesse98}; we take a mean metallicity of
0.3$Z_{\odot}$ and a mean galactic absorption corresponding to $N_H =
3\cdot10^{20}$ cm$^{-2}$. This step provides us with ICM emissivity maps; a further example is presented in Fig. \ref{photconv}.

\begin{figure}
\centering
\centerline{\includegraphics[width=9.5cm]{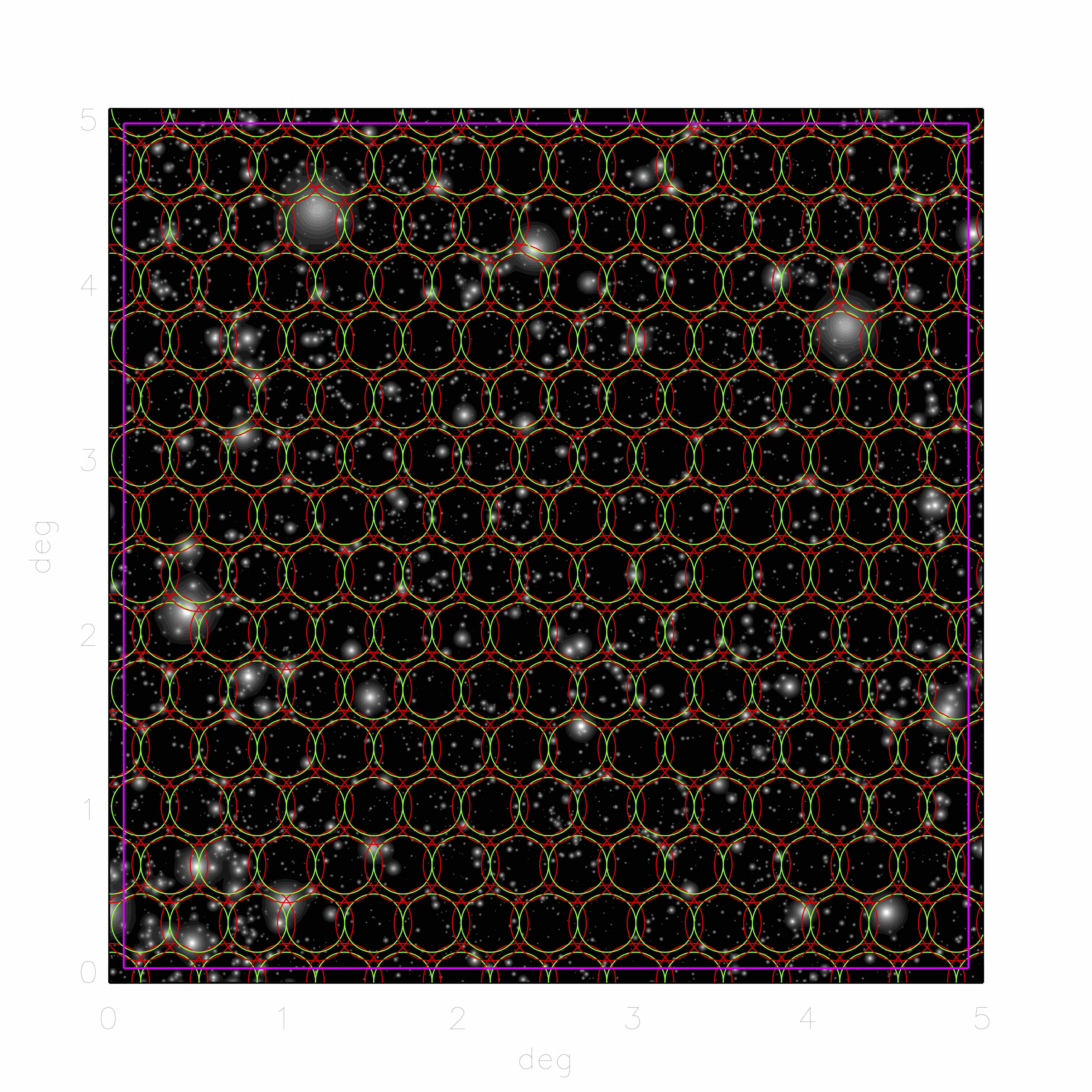}}
\caption{Layout of the XMM pointings over a single 25 \dd\ region; the observations are separated by 10' in RA and Dec. Source detection is
  performed out to a radius of 13' (red circles). For the cosmological
  analysis only sources in the innermost 10' are considered  (green
  circles). To avoid border effects, we discarded all detections
  outside the magenta square.}
\label{tilmap}
\end{figure}

\section{XMM synthetic surveys}

We describe in this section the conversion of the ICM  emissivity maps into XMM images. 
 
\subsection{Survey geometry}
The tiling of a single 25 \dd\ field by XMM observations  is shown in Fig.\ref{tilmap}.  The XMM field of view is 15' but given that the point spread function and the sensitivity are rather poor at large off-axis, we restrict the source detection to  an off-axis of 13'  and consider only  the innermost 10' for the cosmological cluster sample. Moreover, to exclude border effects, we trim all $5\times5$ \dd\ fields off  by 5'. This yields a ``cosmological'' area of 18.22 \dd\ for each subregion, i.e. a total of 710.6 \dd . The number of XMM observations processed in one band reaches $\sim$ 9000. The observations are assumed to be performed with 10 ks exposures and the THIN filter.\\
 
\subsection{Conversion into XMM images}
The Aardvark [0.5-2] keV band maps produced in Sec \ref{photonmap} are 2.5'' images in unit of photons/s/cm$^2$. 
To convolve with the XMM spectral response and effective area, we assume a mean photon energy of 1 keV for all photons: pixel physical fluxes are transformed into XMM count-rate unit. The PSF distortions as well as the vignetting are then applied. This is done separately for the 3 XMM detectors, each with its own specific energy and spatial response to yield, in the end,  event lists as for real observations.

\noindent

\begin{figure}
\centerline{\includegraphics[width=9.5cm]{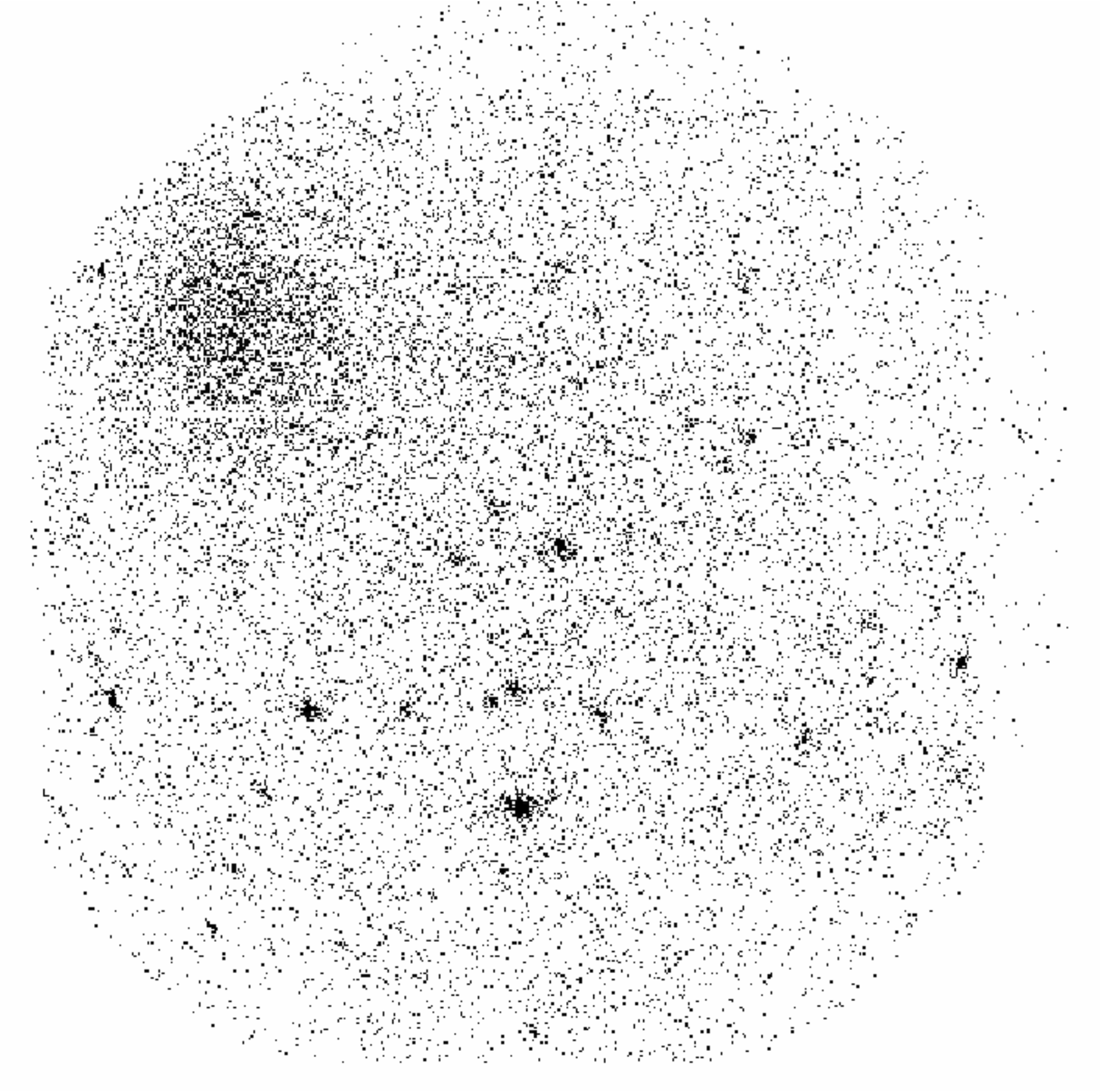}}
\caption{Simulated XMM image (MOS1+MOS2+PN, 2.5'' pixel) obtained for a 10 ks exposure on the region   displayed in Fig. \ref{photconv}. The AGN population  and the diffuse background components  are added to the ICM emission modelled from the Aardvark simulation. All instrumental effects such as the detector spectral responses, the vignetting function and the PSF are taken into account.} 
\label{finalev}
\end{figure}  

\begin{table}[t]
\centerline{\renewcommand{\arraystretch}{1.2}\begin{tabular}{c ||c }
band [0.5 - 2] keV& MOS1+MOS2+PN\\
\hline
\hline
diffuse background *& 5.1 $\cdot$ 10$^{-6}$ cts/s/pix \\
soft-proton *& 2.6 $\cdot$ 10$^{-6}$ cts/s/pix \\
particle background& 2.4 $\cdot$ 10$^{-6}$ cts/s/pix \\
\hline
\end{tabular}}
\caption{Background components added to the  ICM event-list for the  [0.5 - 2] keV band, in which the cluster detection is performed. The pixel size is 2.5". The * indicates the components affected by the instrumental vignetting.}
\label{backs3}
\end{table}
\noindent
\subsection{Back- and foreground photons}
In order to produce most realistic XMM images, the event lists obtained from the ICM are merged with those coming from other source of emission, namely: fore- and background AGN and the various components of the diffuse X-ray background. This latter contribution is summarized in Table \ref{backs3}.
The adopted mean particle background is an average of XMM observations obtained with the closed filter. The diffuse and soft proton backgrounds follow the model proposed by  \citep{snowden08}.  The X-ray AGN population is taken from  the \textit{$\log$N-$\log$S} by \cite{moretti03}, down to a flux limit of 10$^{-16}$ erg/cm$^{2}$/s. AGN are randomly distributed over the XMM field of view, ignoring in the present paper, their spatial correlation and the fact that AGN may be present in cluster centres. The AGN pointlike sources are convolved with the same instrumental effects (energy response, PSF, vignetting) as for the ICM diffuse emission.  Fig. \ref{finalev} shows an example of a final simulated XMM image.

\begin{table}[t]
\centerline{\renewcommand{\arraystretch}{1.2}
\begin{tabular}{c || c c }
\multicolumn{3}{c}{C1 catalogues}\\
\hline
 &\multicolumn{2}{c}{innermost 10'} \\
 & B0& B0.5  \\
\hline
\hline
halo  & 4 483 & 4 273  \\
ambiguous  & 101 & 84  \\
\hline
AGN  & 65 & 72\\
false  & 218 & 214 \\
\hline
Total & 4 867 & 4 643  \\
\hline
\hline
contamination  & 5.8\% & 6.2\% \\ 
density  &6.8/6.3 &6.5/6.1 \\
\hline
\end{tabular}} 
\caption{C1 sources correlated with the input  Aardvark halo and AGN  catalogues. We display the results for the two adopted cluster profiles. Contamination is defined as (AGN+false)/(ambiguous+halo). Densities are computed for both the total and ambiguous+halo detections over over 711 deg$^2$.}
\label{corr13}
\end{table}

\section{Creation of the C1 cluster cosmological catalogues}

The synthetic observations are processed with the {\sc Xamin} pipeline in the same way as real standard XXL observations \citep[e.g.][]{pacaud06, pierre16}. We extract the C1 cluster candidates from the pipeline output lists.  More than 4500 clusters were detected for realisations B0 and B0.5.

\subsection{Correlation with the input catalogues}
\label{correl}
For real observations, the {\sc Xamin} pipeline is used only at the cluster detection stage on the individual XMM observations: [1] source detection is routinely perform within the innermost 13' of the detector but we usually restrict the cosmological sample to the inner 10', the radius at which the sensitivity reaches 50\% of the on-axis value \citep{clerc12a}; [2] measurements of cluster properties are subsequently performed in a semi-interactive mode \citep[e.g.][]{giles16} to cope in an optimal way with the particularity of each source e.g. AGN contamination, local background removal and possible irregular cluster shapes. This is an important step since the quality of the cosmological analysis heavily relies on the precision of these  measurements.\\
For the present test-study based on simulations, it was not conceivable to measure in this way some $2 \times 4500$ objects. We thus correlated the pipeline output catalogues with the input simulated catalogues containing the cluster mass, luminosity, temperature and core radius information. In this way, we were able to assign to each detected C1,  total XMM countrates in the chosen bands, following the same principles as described for the production of the [0.5-2] keV count-rate map.
As in previous studies \citep{pacaud06}, we use a 37.5'' radius for the correlation with the Aardvark cluster list and a 6'' radius for the random AGN list. The correlation outputs were flagged as follows:
\begin{itemize}
\item Cluster : when a C1 source  is matched to an input Aardvark halo
 \item AGN: when C1 source is matched to an input AGN (rare case)
\item ambiguous: when the two previous conditions are both true
\item false: when none of the previous conditions is true. 
\end{itemize}
The results of the correlation are reported in Tab. \ref{corr13}. They show a somewhat higher C1 contamination rate than reported in our previous analytical simulations \citep{pacaud06}. The $\sim 5 \%$ fraction of fake sources can be explained by the fact the analytical simulations avoided cluster overlap, while projection effects naturally occur when using a cosmological light-cone, creating multiple cluster detections for some peculiar lines of sight. In the real observation regime, the C1 catalogue is systematically screened by two independent persons to remove obvious fake detections.

\subsection{The cosmological sample}
By restricting the cluster catalogue to the inner 10', we expect a higher S/N for the detected sources and a better positional accuracy. We exclude from the cosmological sample all sources flagged as AGN and fake. We subsequently define the C1 CLEAN sample as the C1 sources flagged as cluster and ambiguous within 10' and consider this sub-sample as the best trade-off between a fully automated procedure and a dedicated interactive screening.
The corresponding CLEAN survey area amounts to  710.56 deg$^2$   with XMM. The C1 density is 6.3/deg$^2$ is 6.1/deg$^2$ for the B0 and B0.5 configurations respectively.

\subsection{Measurement errors}
\label{errmeasure}
\begin{figure}
\centerline{\includegraphics[width=9.5cm]{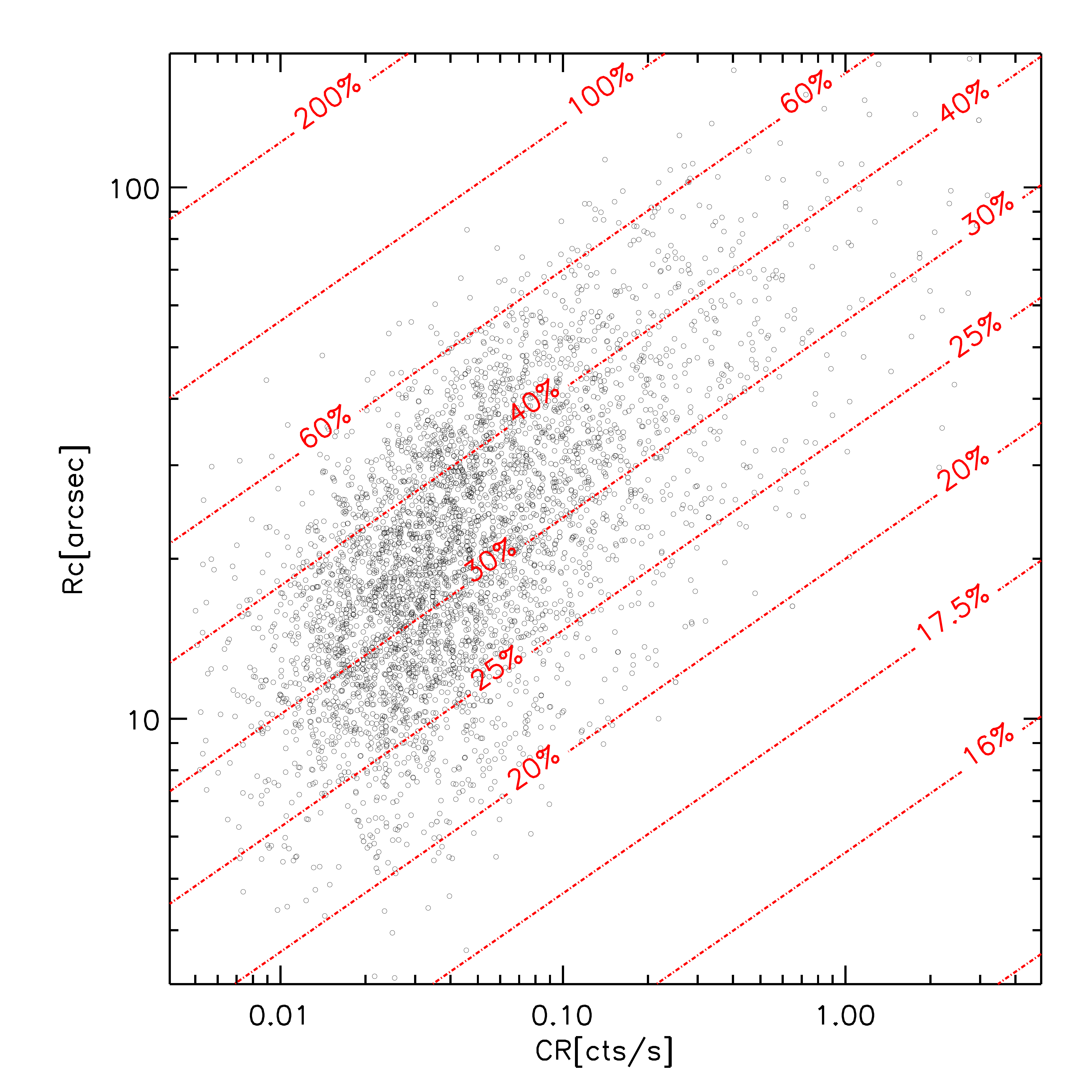}}
\caption{The red lines show the adopted measurement-error model as a function of the nominal total [0.5-2] keV count-rate and apparent core radius; the black circles  are the detected Aardvark C1 clusters, drawn to highlight the cluster locus in this parameter space. Practically, the error on CR and $r_{c}$ are randomly ascribed from a log-normal distribution with the dispersion given in the plot. Errors on HR$_{1}$ and HR$_{2}$ are assumed to be respectively the same   and the double values obtained for a given [CR, $r_{c}$] combination. The model assumes a mean vignetting value.} 
\label{diagonal}
\end{figure}

\begin{figure}
\centerline{\includegraphics[width=9.5cm]{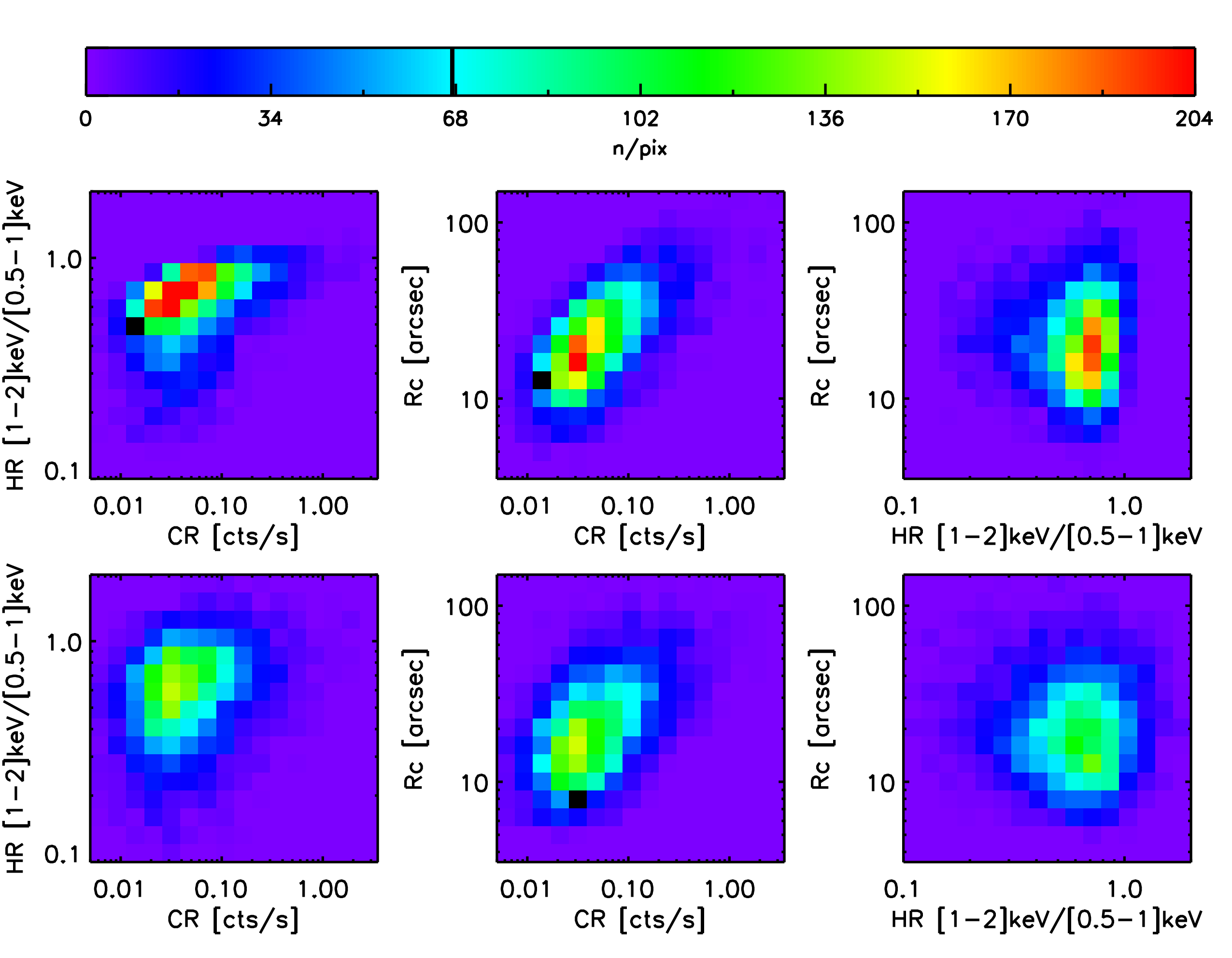}}
\caption{Effects of measurement errors on the C1 CLEAN sample. The plots show from left to right the 2D diagrams CR-HR, CR-$r_{c}$ and HR-$r_{c}$. The first row stands for the nominal CR, HR, $r_{c}$ values stored in the Aardvark catalogues. The second row shows the result of the implementation of the error model displayed in Fig. \ref{diagonal}.} 
\label{erronaa}
\end{figure}

Last step is to ascribe realistic error measurements to each cluster parameter entering the XOD. The chosen error model is presented in Fig. \ref{diagonal} and is based on our experience with analytical simulations. It is applied to the total count-rates and to the core radii derived from the Aardvark catalogues (Sec. \ref{correl}). To simplify the formalism, errors are given as a function of the [CR, $r_{c}$] combination (which is also the plane used for the cluster selection) and we assume that they have the same amplitude for CR, HR$_1$ and $r_{c}$, ; for the second color, HR$_2$,  we double this value given the XMM sensitivity drop in the hard band. The effect of the error measurements on the XODs is illustrated in Fig. \ref{erronaa}.

\section{The cosmological fit}

We now describe the cosmological fit on the Aardvark catalogue of X-ray halos, prepared as described in the previous sections. Our basic analysis sticks to the same set of free parameters as in paper III, namely [$\Omega_{m}, ~ \sigma_{8}, ~ x_{c}, w_{0}$], assuming that the scaling relations are known and evolve self-similarly. In a second step, we open as free parameters, the coefficients of the M-T relation.  
The Amoeba cosmological fitting on the XOD is extensively described in paper III and is summarized section \ref{aspix-f}. For the MCMC analysis on the same XOD, we use a  Metropolis-Hastings algorithm \citep{metropolis53}. Parallel chains are considered to have converged by applying the G-R criterion with $r < 1.03$ \citep{gelman92}. The fit results are discussed in Sec. \ref{discussion}.

\subsection{Analysis of the 700 deg$^2$ survey} 
\subsubsection{Testing constraints from the mass distribution alone} 
\label{zm500} 
Fig. \ref{aamasscmp} shows an overall excellent agreement between the measured Aardvark mass function and the Tinker modelling assumed in our XOD fit. A moderate deviation is nevertheless observed in the high-redshift slice above $log(M) \sim  14.5 M_{\odot}$, a range expected to have a high weight in the cosmological analysis. In order to test the impact of this particular uncertainty, we thus first run the cosmological fit on the mass function alone.
For this, we assume (i) an ad-hoc pure mass-selection giving 4296 clusters, which is very comparable to the number of C1 clusters (Table \ref{corr13});  (ii) no error measurements on the masses (in the selection and in the cosmological analysis).

\begin{figure}
\includegraphics[width=9.5cm]{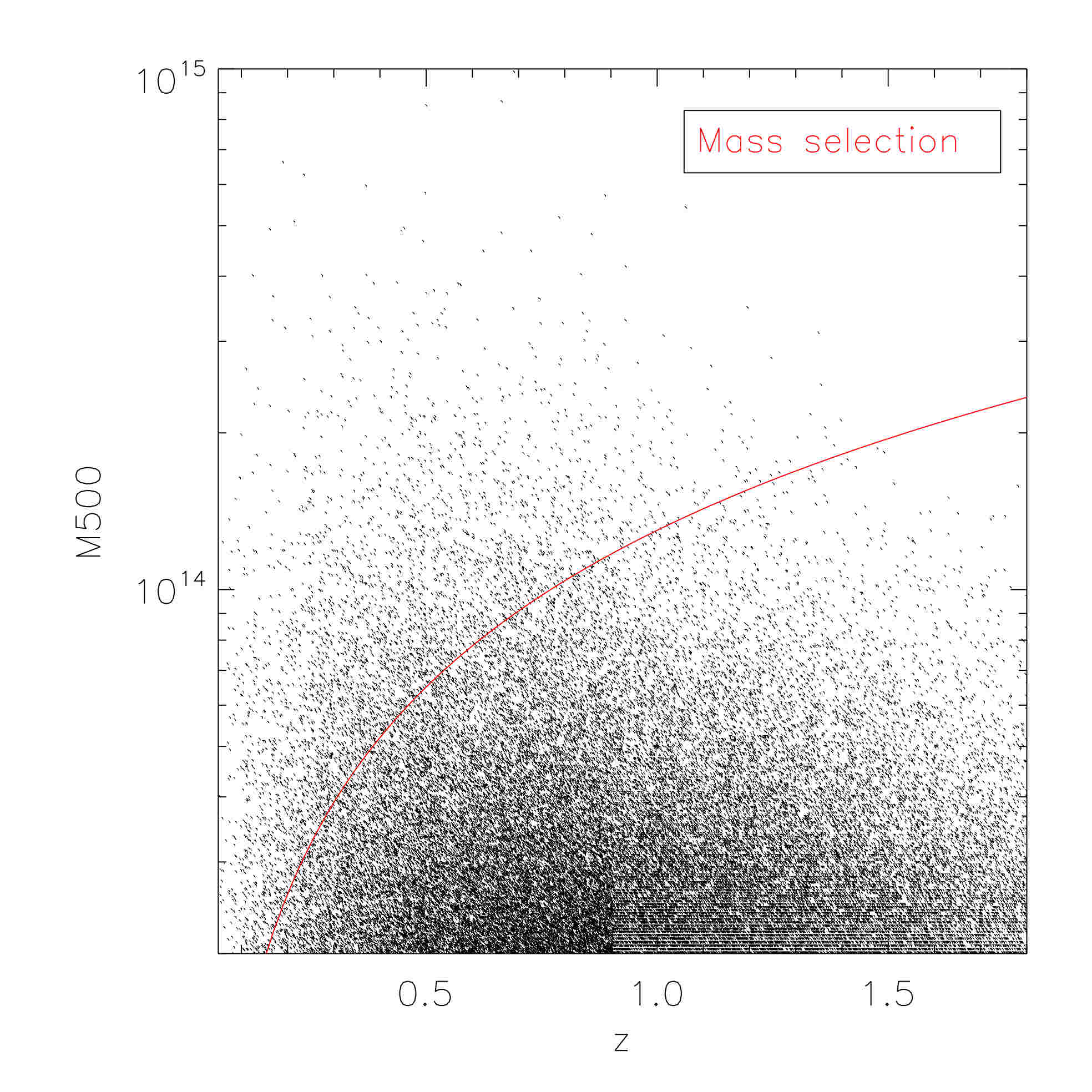}
\caption{X-ray analogous mass  selection used to test the impact of the deviation between the Aardvark and Tinker mass functions.} 
\label{mass-sel}
\end{figure}

The selection is displayed  on  Fig. \ref{mass-sel} and the results, along with the Fisher analysis predictions, are given in Table \ref{mzamoeba}. The cosmological fit for this particular halo catalogue has been performed with Amoeba on the [$M_{500},~z$] distribution  using 100 different starting points, as for the XODs. We  mention at this stage that the $x_{c}$ value is poorly constrained since this parameter does not intervene in any stage of the mass-fit (nor in the selection, neither in the mass-measurements, which are assumed to be perfect).
The conclusion of this  exercise is that the 5 \% discrepancy observed at the high-end of the mass function, between the fitting model and the simulations, has a negligible effect on the cosmological analysis given the size of the assumed measurement uncertainties.

\noindent 
\begin{table*} 
\begin{tabular}{l | c c  c}
\hline
Parameter    &$\Omega_m$ & $\sigma_8$ &$w_0$ \\
fiducial   &0.23 & 0.83  &-1  \\
\hline \hline
Fit of the mass-function (best-10 values) &0.227  & 0.828 & -0.981 \\
Fisher predictions & $\pm$0.0001 & $\pm$0.004 &  $\pm$0.031 \\
\hline
\end{tabular}
\caption{Fit of the mass function (dn/dM/dz) for halos selected as in Fig. \ref{mass-sel}. Uncertainties on the mass measurements are assumed to be null.  
The errors predicted by the Fisher analysis assume that the real (Aardvark) universe and the fitted model have exactly the same mass function (namely Tinker's), hence indicate the shot noise level for a 700 \dd\ area.
The comparison between the Fisher predictions and the fit results provides an estimate of the impact of the Tinker hypothesis for this particular halo sample.}
\label{mzamoeba}
\end{table*}
\noindent

 \subsubsection{Signal variable diagrams}
 
\begin{figure}
\centering
\centerline{\includegraphics[width=9.5cm]{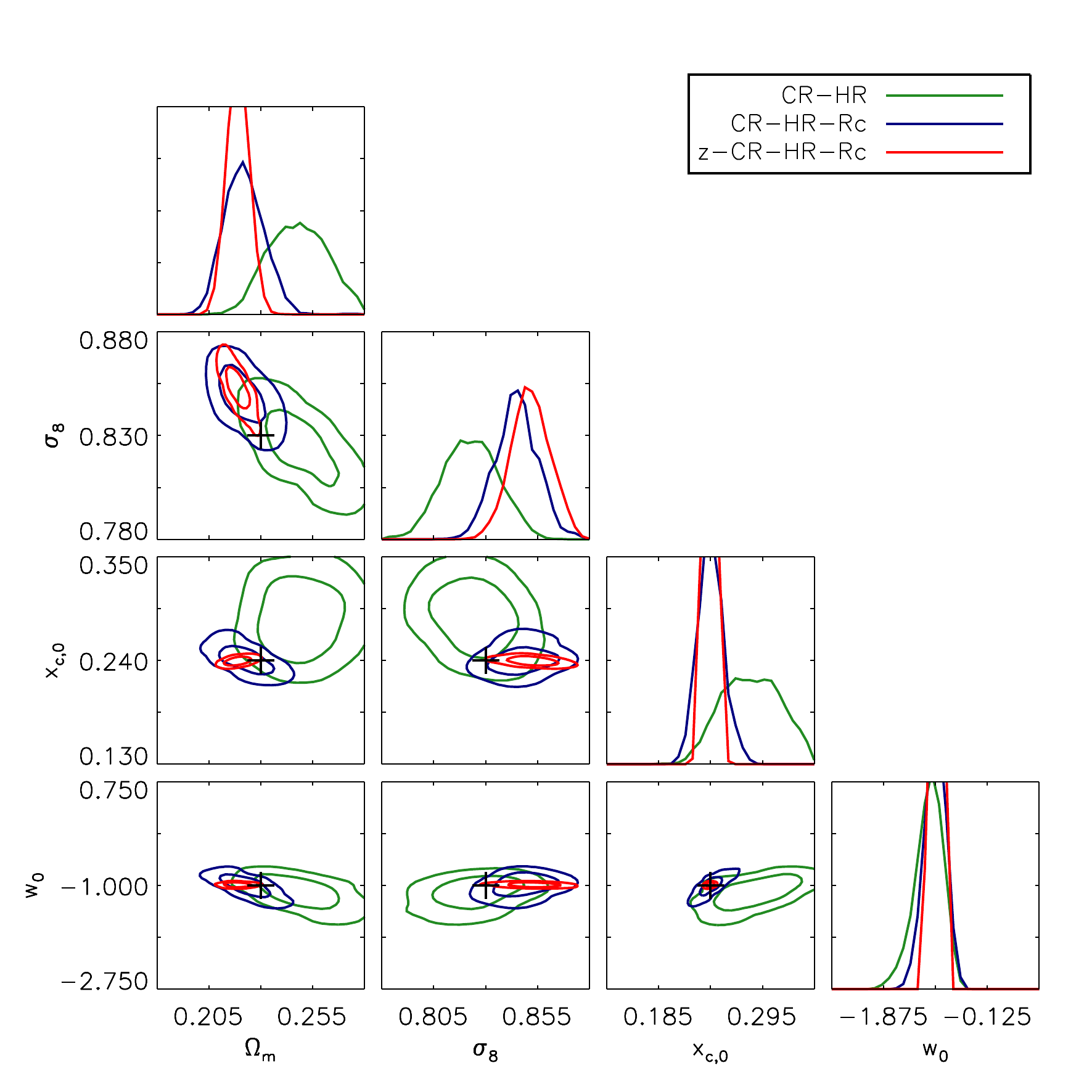}}
\caption{Confidence regions at the 68\% and 95\% levels and 1D marginalized
  distribution for the studied parameter subset ($\Omega_m$, $\sigma_8$,
  $x_{c}$, $w_0$). The cross indicates the fiducial model. The MCMC analysis was run on an effective sky area of 711 deg$^2$ for the CLEAN C1 catalogue, involving some 4300 clusters. Fit for $z-$CR-HR-$r_c$ is in red, for
CR-HR-$r_c$ is in blue and for CR-HR in green.}
\label{conf1}
\end{figure}

\begin{figure}
\centering
\centerline{\includegraphics[width=9.5cm]{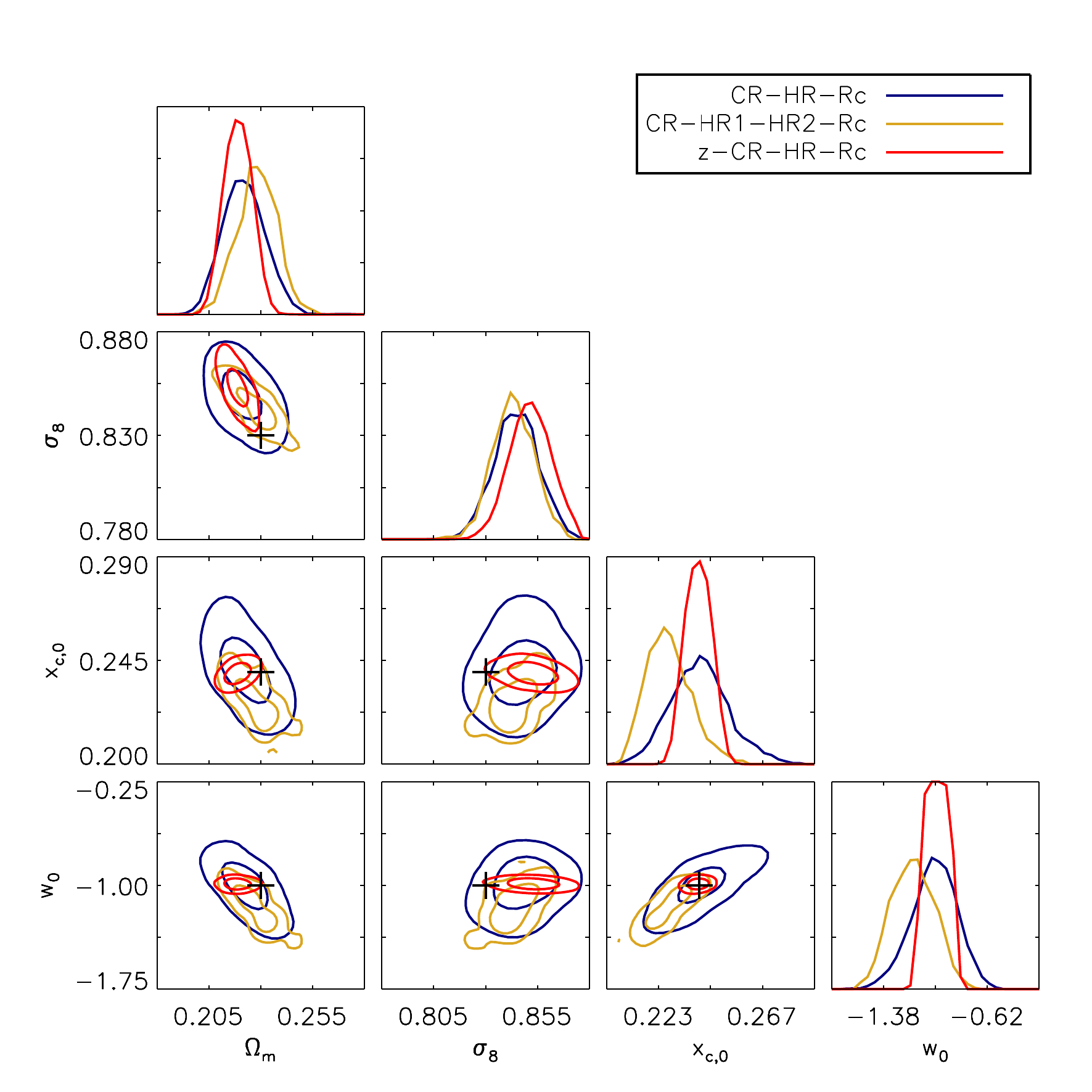}}
\caption{Same as Fig. \ref{conf1}, for other observable combinations. This  figure allows a visual comparison of the relative constraining power of $z$ and HR$_2$.}
\label{conf2}
\end{figure}

We now turn to the cosmological analysis of the C1 "CLEAN" Aardvark catalogue and test combinations involving an increasing number of  signal variables : CR-HR$_1$, CR-HR$_1$-$r_{c}$, z-CR-HR$_1$-$r_{c}$, CR-HR$_1$-HR$_2$-$r_{c}$. All diagrams include scatter in the scaling relations (L, T, R$_{c}$) and error measurements as described above.
Each XOD diagram is fitted using, either a MCMC method (providing uncertainties) or  100 Amoeba runs. Given that the Amoeba route does not provide errors on the output parameters, we estimate them by averaging the results of 10 X 700 \dd\ analytical toy-catalogues, following the methodology introduced in paper III. We also provide the predictions from the Fisher analysis. The results are gathered in Table \ref{710mcmc}. The graphic representation of the MCMC output is displayed in Fig.  \ref{conf1} and \ref{conf2}.

\begin{table*}[t]
\centerline{\renewcommand{\arraystretch}{1.3}
\begin{tabular}{c c l| c c |c c}
\hline
ID &Observable combination & Fitted parameters & $<p>$ & \em{best-10} & Toy catalogues[x10]& Fisher\\
& & & MCMC & Amoeba & Amoeba & analysis\\
\hline
\hline
A1&CR-HR$_1$ &$\Omega_m$ & 0.249$^{+0.014}_{-0.019}$  & 0.245& 0.234$\pm$0.019 &0.23 $\pm$ 0.013  \\
  & &$\sigma_8$ & 0.823$\pm$0.014  &0.825 & 0.830$\pm$0.018& 0.83 $\pm$ 0.012  \\
 & &$x_{c,0}$& 0.285$^{+0.033}_{-0.034}$  & 0.290 & 0.232$\pm$0.024& 0.24 $\pm$ 0.031 \\
 & &$w_0$ & -1.117$^{+0.212}_{-0.218}$  & -1.037 & -1.204$\pm$0.296& -1.00 $\pm$ 0.246\\
\hline
A2&CR-HR$_1$-$r_c$ &$\Omega_m$ & 0.222$\pm$0.010 & 0.220 & 0.226$\pm$0.013 & 0.23 $\pm$0.012 \\
 & &$\sigma_8$ & 0.846$^{+0.011}_{-0.010}$ &0.846 &  0.832$\pm$0.015& 0.83 $\pm$ 0.011 \\
 & &$x_{c,0}$& 0.240$^{+0.011}_{-0.013}$ & 0.247 & 0.248$\pm$0.014 & 0.24 $\pm$  0.017\\
 & &$w_0$ & -1.009$^{+0.153}_{-0.144}$  & -0.969 & -0.980$\pm$0.198 & -1.00 $\pm$ 0.21\\
\hline
A3& CR-HR$_1$-$r_c$-z &$\Omega_m$ &0.219 $\pm 0.005$  & 0.218 &0.229$\pm$0.004 & 0.23 $\pm$ 0.005\\
 & &$\sigma_8$ & 0.852$\pm 0.009$ &0.854 &0.832$\pm$0.009 & 0.83 $\pm$ 0.009   \\
 & &$x_{c,0}$& 0.240$\pm $0.003 & 0.239 &0.240$\pm$0.003& 0.24 $\pm$ 0.003 \\
 & &$w_0$ & -0.990$^{+0.029}_{-0.027}$  & -0.990 &-1.041 $\pm$0.033& -1.00 $\pm$ 0.032 \\
\hline
A4&CR-HR$_1$-HR$_2$-$r_c$ &$\Omega_m$ &0.228$^{+0.008}_{-0.009}$  & 0.227 &0.226$\pm$ 0.013 & 0.23 $\pm$ 0.008 \\
 & &$\sigma_8$ & 0.844$^{+0.008}_{-0.009}$ &0.843 &0.833 $\pm$ 0.012 & 0.83 $\pm$ 0.010  \\
 & &$x_{c,0}$& 0.226$^{+0.008}_{-0.009}$ & 0.229 &0.247 $\pm$ 0.012 & 0.24 $\pm$ 0.009 \\
 & &$w_0$ & -1.166$^{+0.148}_{-0.146}$ & -1.121 & -0.975 $\pm$ 0.195 & -1.00 $\pm$  0.113\\
\hline
\end{tabular}}
\caption{ Summary table for the cosmological analysis of the Aardvark C1 CLEAN catalogue over 711 deg$^2$. The first column gives the run ID. The second column lists the signal variables used in the fit and the third one, the subset of free parameters. The fourth and fifth columns show the results from the  MCMC analysis at the 68\% confidence level and from the Amoeba best-10 fit, respectively.  The sixth column shows the results obtained by running Amoeba over 10 toy catalogues of 700 deg$^2$, for which the mass function is taken to be Tinker's. The last column shows  the  Fisher analysis forecast for $1\sigma$ errors .}
\label{710mcmc}
\end{table*}

\noindent

\subsubsection{Scaling relation evolution}

We also investigated the behaviour of ASpiX in the case where the parameters of the M-T relation are totally unconstrained. In this configuration we switch from 4 to 9 free parameters. The results are reported in Table \ref{sclfit}.

\begin{table}[h]
\begin{tabular}{l | c c c }
\hline \hline
 Parameter &  MCMC fit & Amoeba {\em best-10} & Fisher analysis\\  
$\Omega_m$ &  0.228 $\pm$ 0.020 & 0.207 & 0.23 $\pm$ 0.025 \\
$\sigma_8$ &  0.876 $\pm$ 0.073 & 0.814 & 0.83 $\pm$ 0.156 \\
$w_0$ & -0.981 $\pm$ 0.053 &-0.940 & -1.00 $\pm$ 0.065 \\
$x_{c}$&  0.249 $\pm$ 0.016 &0.258 &  0.24 $\pm$ 0.034\\
$\sigma_{x_{c}}$&  0.500 $\pm$ 0.019 & 0.504& 0.50 $\pm$ 0.023 \\
$\alpha_{MT}$ & 1.538 $\pm$  0.096 &1.453 & 1.49 $\pm$ 0.169 \\
$\gamma_{MT}$ & 0.268 $\pm$ 0.136 &0.162 & 0.00 $\pm$ 0.244 \\
$C^{MT}$ &  0.502 $\pm$ 0.140 &0.490 & 0.46 $\pm$ 0.297 \\
$\sigma_{MT}$ &  0.258 $\pm$ 0.133 &0.112 & 0.10 $\pm$ 0.206 \\
\hline \hline
\end{tabular}
\caption{Fit results ($z-$CR-HR-r$_c$) over the 711 \dd\ Aardvark C1 CLEAN catalogue when  cosmological and cluster physics parameters are let free.}
\label{sclfit}
\end{table}

\subsection{Analysis for the 39 $\times$ 18.22 deg$^2$ surveys}
\label{cosmosmall}
\begin{figure}[t]
\centering
\centerline{\includegraphics[width=9cm]{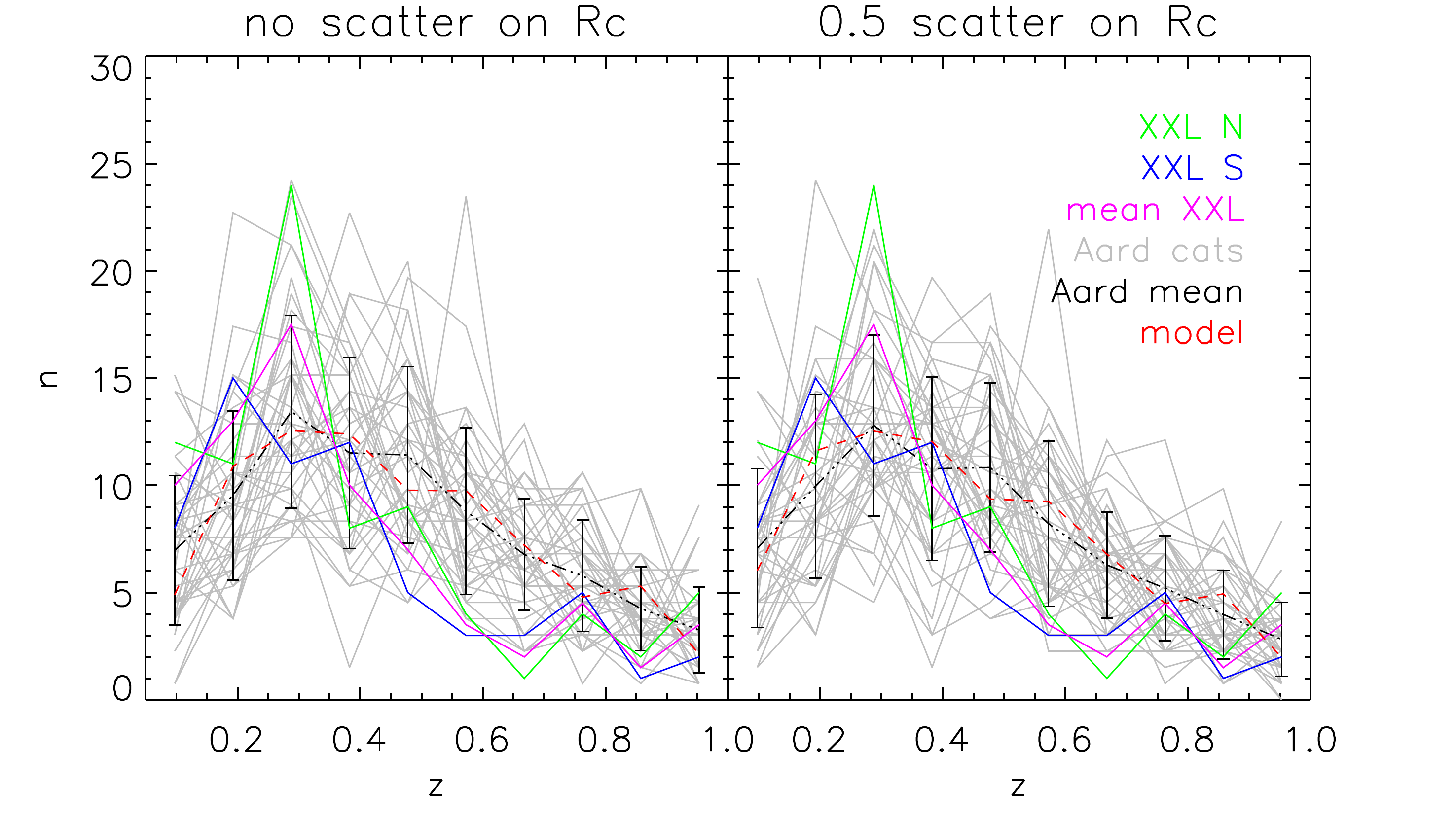}}
\caption{Redshift distribution of the detected C1 Aardvark clusters 
 for the for B0 (left) and B0.5 (right) profile configurations. Gray lines show the cluster selected population and correspond each to 18.22 \dd\ map. The black-dash dotted line stands for the
mean and the error bars show the 1-$\sigma$ deviation. The red-dash line shows our fiducial model (X-ray mapping of the halos + analytical selection). All distributions are normalized to 13.8 deg$^2$ to
match the effective area of the XXL Northern (green solid) and XXL Southern
(blue solid) fields considering only the pointing innermost 10' (XXL paper XX, Adami et al submitted). The mean of the two XXL fields is in magenta.}
\label{zaa}
\end{figure}

In a second step, we investigate the constraining power of the 18.22 deg$^2$ individual maps. 
This is of particular practical interest since these represent approximately the coverage of one XXL survey field, when considering the innermost 10' of the XMM detector. In Fig.  \ref{zaa}, we compare the redshift distribution of the 39 Aardvark sub-fields with that of XXL. 
While the many parameters of our Aardvark modelling ought not to be totally matching  reality as viewed by both XXL fields (cosmology, scaling relations, XMM background), the overall shapes of the redshift distributions appear to be very compatible.

We tested the CR-HR$_1$-$r_c$-z and CR-HR$_1$-HR$_2$-$r_c$ XOD by applying the Amoeba fitting on our  set of free parameters.
The results presented in Table \ref{mapscosmo}  summarise the 3900 fits (100 for each catalogue);  errors are approximated by the 1-$\sigma$ deviation from the mean of the 39  averaged best-10 fits obtained from each sub-catalogue. For comparison we show the predictions from the Fisher analysis.

\begin{table*}[ht]
\centerline{\renewcommand{\arraystretch}{1.2}
\begin{tabular}{c l l | c c c c}
ID& parameter & &$\Omega_m$ & $\sigma_8$ & $x_{c,0}$&$w_0$ \\
\hline
S5& \textbf{cat x 39}  & CR-HR$_1$-$r_c$-z &  &&&\\
&$<$ $p_{best-10}$ $>$ Amoeba & &0.222 $\pm $0.046  & 0.857$\pm $0.080  &  0.240$\pm $0.024  &  -1.022$\pm $0.208  \\
\hline
S6 & \textbf{cat x 39}  & CR-HR$_1$-HR$_2$-$r_c$ &  &&&\\
&$<$ $p_{best10}$ $>$ Amoeba & &0.222 $\pm $0.055  & 0.855$\pm $0.075  &  0.246$\pm $0.041&-1.103$\pm $0.506  \\
\hline
\hline
&Fisher &CR-HR$_1$-$r_c$-z  &0.23 $\pm $ 0.031 & 0.83$\pm $0.060&0.24$\pm$ 0.021&-1.00$\pm$0.175\\
&analysis& CR-HR$_1$-HR$_2$-$r_c$ &0.23 $\pm $0.050 &0.83 $\pm $0.063 &0.24$\pm$0.051 &-1.00$\pm$ 0.705\\
\end{tabular}}
\caption{Cosmological analysis performed on the Aardvark C1 ``CLEAN'' samples corresponding to  the 39 $\times$ 18.22 deg$^2$ sub-maps (effective area).  Two XOD are considered: CR-HR$_1$-$r_c$-z  and CR-HR$_1$-HR$_2$-$r_c$.   The displayed statistics are the average of  the best-10 values obtained for each of the 39  small fields and associated standard deviation. The last two rows give the predictions from the Fisher analysis}
\label{mapscosmo}
\end{table*}

\begin{table}[t]
\centerline{\renewcommand{\arraystretch}{1.6}
\begin{tabular}{c || c c c }
  & \multicolumn{3}{c}{$\sigma_{\ln R_{c} | R_{500c}}$}\\
\cline{2-4}
$x_{c}$& - & 0.25 & 0.5 \\
\hline
\hline
0.1 & 7.9/deg$^2$ & 7.7/deg$^2$ & 6.7 /deg$^2$\\
0.24 & 6.3/deg$^2$ & 6.2/deg$^2$ & 6.1/deg$^2$\\
0.4 & 3.2/deg$^2$ & 3.5/deg$^2$ & 4.0/deg$^2$\\
\hline
\end{tabular}}
\caption{C1 cluster density (analytical calculations) as a function of cluster intrinsic size ($R_{c}$) and scatter in the $M-R_{c}$ relation.
  The adopted cosmology and X-ray cluster scaling relations are given in Sec. \ref{simulations} and the selection function is displayed in Fig.  \ref{selfunc}.}
\label{ncountcmp}
\end{table}

\section{Discussion}
\label{discussion}
The purpose of the present article is to quantify the behaviour of the ASpiX method in more realistic conditions than the preliminary study presented in paper III,  based on analytical toy-catalogues. Here, the use of template-based simulations, transformed into real-sky XMM images, allowed us to implement the effect of the selection function as well as a more realistic error model for the considered variables. We globally confirm the very positive results of paper III and discuss our findings below.

\subsection{Error estimates on the cosmological parameters - 711 \dd\ catalogue}
Table \ref{710mcmc} summarizes the main outcome of the study, where we compare for the 711 \dd\ survey, (i) the effect of adding signal variables, (ii) the errors returned by the MCMC and Amoeba fitting methods and (iii) the prediction of the Fisher analysis (FA). At this stage, we assume that the cluster scaling relations are perfectly known.\\
Logically, considering successively CR-HR, CR-HR-$r_{c}$ and  CR-HR-$r_{c}$-z, the uncertainties  decrease when the number of dimensions describing the cluster population increases. We recall here that the errors from the Amoeba fitting are quantified by running numerous ($\geq 10$) independent realisations of simulated catalogues. Given that only one 711 \dd\ realisation was available, we analytically created ten 700 \dd\ `toy-catalogues' : the differences with respect to the 711 \dd\ simulation is that the objects (1) were created exactly following the Tinker mass function and (2) were not selected `in situ' by the {\sc Xamin} pipeline but using the analytical selection function (Fig. \ref{selfunc}), the same that is used for fitting the XOD and performing the FA; but the error model on the observables is the same. \\
All numbers recorded in Table \ref{710mcmc} are displayed with 3 decimal digits for the purpose of comparison, but this should not be ascribed a high significance, since not all systematic effects have been considered in the error budget. All in all, the three approaches deliver very comparable error estimates,  somewhat larger for Amoeba, hence better bracketing the fiducial cosmological model. Interestingly, the fit of the N(M, z) distribution, assuming no error on mass and a perfect selection function (Table \ref{mzamoeba}) does not appear to produce better results than CR-HR-$r_{c}$-z in real-sky conditions for $w_{0}$ - the FA predictions are indeed at the same level. Table \ref{710mcmc} suggests that, with our current working hypotheses, any deviation  between the Aardvark mass function with respect to Tinker's has a negligible impact on our results. \\
Finally, we compared the efficiency of adding a forth dimension to the XOD: either as redshift (run A3) or as a second X-ray color (HR$_2$, run A4). Although both induce a significant improvement with respect to   CR-HR$_1$-$r_c$, the redshift information appears to outperform the  colour information (as inferred from the error bars). This is easily understandable since only the knowledge of  redshift breaks the temperature-redshift degeneracy in a pure Bremsstrahlung spectrum.   The addition of a second colour solely brings a second measurement (hence refining the first one) of the $T/(1+z)$ degeneracy (cf bottom central panel of Fig. \ref{XODex}). Of course, the spectra considered here do contain emission lines from metals (APEC plasma code) but given the small number of collected photons, the effect on the degeneracy is small.\\
We note that the particular Aardwark realisation seems to converge (when z is available) to a point that is beyond the 1-$\sigma$ error for $\sigma_8$ and $\Omega_m$, but perfect for $w_0$ and $x_c$; this is not unexpected from the statistical point of view. Indeed, among the 10$\times$ 700 \dd\ toy-models   generated for this study, we also found two catalogues yielding somewhat displaced values: $\Omega_m =0.223,~ \sigma_8=0.842, ~x_c=0.240, ~w_0=-1.036$.  

\subsection{Error estimates on the cosmological parameters - 39 $\times$ 18 \dd\ catalogues}
Another way to scrutinize the ASpiX output is to apply the method individually on the 39 $\times$ 18.22 \dd\ sub-regions whose  assembly constitutes the 711 \dd\ area.  By averaging the Amoeba fitting of each XOD, we obtain the mean uncertainty on the cosmological parameters expected for a  18.22 \dd\ area. The results   are given in Table \ref{mapscosmo}. The mean values are well within the 1-$\sigma$ expectations. These error estimates are comparable to the Fisher predictions, but do not exactly follow the expected SQRT(area) scaling as can be inferred from Table \ref{710mcmc}. It is likely that with such a small area (some 110 clusters in average per field)  the sampling of the XOD  in its  four dimensions has to be revisited, in order to optimise the fitting procedure. We defer this question to a future paper.

\subsection{Fitting cosmology along with cluster scaling relations}
In paper III, we showed that for large enough surveys it possible to fit at the same time, and to recover with excellent accuracy,  the cosmological parameters and the coefficients of both cluster scaling relations (scatters were assumed to be known). This was demonstrated assuming a 10 000 \dd\ area and using the Amoeba fitting. Here, we show in Table \ref{sclfit} the results of the MCMC fit  for the 711 \dd\ Aardvark catalogue when the coefficients of the M-T relations are let free, as well as the scatters of the M-Rc and M-T relations. While the cosmological parameters, the M-Rc relation and the slope of the M-T relation can be recovered at a level better than 10\%, we observe larger errors for the amplitude, evolution and   scatter of M-T. This is not surprising as the scatter has a strong effect on the cluster selection selection process, hence induces additional degeneracy in the cosmological analysis  \citep{pacaud06,allen11}. 
In this run, the coefficients of the L-T relation are  held fixed. In practice, the L-T relation is the easiest cluster scaling relation to determine and can be easily computed for a 10ks cluster sample \citep[e.g.][]{giles16}; it can then be plugged as a prior into the cosmological analysis \citep[e.g.][]{pacaud16}.\\
In average, the MCMC errors are smaller than predicted by the FA (and do not always bracket the input fiducial values). We explain this by the fact that, so far, we do not allow for uncertainties in the selection function. When both cluster physics and cosmological parameters are let free, the MCMC may converge to a peculiar solution (close to the fiducial one, but with a higher likelihood); being forced to consider the cluster catalogue as perfect, the MCMC may ascribe too small errors to its preferred solution.

Finally, we investigated in more detail the effect of the scatter in the M-Rc relation on the number of detected clusters. Table \ref{ncountcmp} summarises our calculations, still assuming the C1 selection function. The results indicate that the detection rate depends in a non-intuitive manner on the cluster size and  scatter : for $0.1< x_c < 0.24$, the more peaked the clusters and the smaller the dispersion, the larger the number of detected clusters; for $x_c = 0.4 $,  the smaller the dispersion, the smaller the detection number. Hence, for the range of  $x_c$ values usually postulated ($x_c = 0.1-0.24$),  the scatter of the M-Rc relation has the opposite effect as that of the L-T relation.

\section{Summary and conclusion}

This article presents an in-depth formal analysis of ASpiX, an observable-based method for the cosmological analysis of X-ray cluster surveys. The basic working hypothesis is that only shallow survey data are available, which enable the measurements of cluster count-rates, hardness-ratios and apparent sizes. The method allows the inclusion of all detected clusters and combines in a single fitting procedure the cosmological parameters, the cluster scaling relations and the survey selection function.
The tests are performed on a  711 \dd\ semi-analytical simulation (Aardvark). The perfect  X-ray emissivity sky map associated to the dark matter halos is in turn converted into XMM event lists, using a  state-of-the-art procedure that reproduces all observational effects. Clusters of galaxies are then detected and selected using the {\sc Xamin} XXL pipeline. \\
The main upgrades with respect to paper III based on analytical toy-catalogues, is the in situ  selection  function as well as a more realistic modelling of the measurement errors for the considered variables. We moreover complement our simple minimisation  routine (Amoeba)  by the use of an MCMC code. The uncertainties  quoted throughout the paper are given for comparison purposes and should not be considered as final, since a number of second order systematics were not considered. \\
We confirm and extend the results of paper III, namely that the method is as reliable as the approach based on cluster counts as a function of mass and redshift. The method is  modular and flexible in the sense that, in practice, there is no need to  re-measure the cluster parameters for each tested cosmology (e.g. $ M = {\bf F}[L_x], ~ L_x= {\bf G}[D_l, R_{500}^{proj}], ~ R_{500}^{proj} = {\bf H}[M,D_a]$). The number of parameters (cosmology and physics) that can be simultaneously and efficiently fitted depends, as for any approach, on the number of clusters available for the  analysis.  The MCMC fit tends to give smaller error bars than the error estimates obtained by applying Amoeba on toy-catalogues, but the latter are in better agreement with the FA predictions.  The Amoeba fitting has the advantage of being some 4 times faster than the MCMC (e.g. for run A2, running 100 Amoeba fits on the 100 CPUs takes 3 hours; 10 additional toy-catalogues for the error calculation require 30 hours; the MCMC takes 6 days).\\
Last step before applying the method on real observations, will consist in extensive tests on hydrodynamical simulations. This will allow us to quantify the effect of cluster irregular shapes (on the selection function and on the measurement of the cluster properties) and that of central AGNs  on the final error budget. In particular, we have developed a formalism to implement the X-ray AGN properties in hydrodynamical simulations \citep{koulouridis17}, which will replace our current random modelling of the AGN population\footnote{Our preliminary findings indicate that the presence of a central AGN in a cluster can modify in either way the C1/C2 ranking of that cluster, depending on the AGN to cluster flux ratio and cluster apparent size}.   Since the X-ray cluster properties, especially for objects below $10^{14} ~ M_\odot$, are affected by non-gravitational physics, we shall derive selection functions for various plausible feedback models;  this will further allow us to evaluate the uncertainties on the selection function. Errors on the cosmological parameters will be estimated by enlarging the `toy-catalogue' set, aiming at, at least, 20 realisations directly drawn from the hydrodynamical simulations. We shall quantify the systematics and covariance between the various parameters, issues that we have not considered so far. One further point regards the sampling of the XOD, which will have to be optimised depending on the number of detected clusters. We shall then be in a position to perform a fully consistent error analysis as a function of  ICM physics, survey depths and background levels.

\begin{acknowledgements}
We  are grateful to Christophe Adami, Dominique Eckert, Elias Koulouridis, Amandine Le Brun, Ian McCarthy and Jean-Baptiste Melin for useful discussions 
\end{acknowledgements}

\bibliographystyle{aa} 
\bibliography{../../../mmplib}{}
\end{document}